# A Survey of Limitations and Enhancements of the IPv6 Routing Protocol for Low-power and Lossy Networks: A Focus on Core Operations

Baraq Ghaleb, *Student Member, IEEE*, Ahmed Al-Dubai, *Senior Member, IEEE*, Elias Ekonomou, Ayoub Alsarhan, Youssef Nasser, *Senior Member, IEEE*, Lewis Mackenzie, *Member, IEEE* and Azzedine Boukerche, *Fellow, IEEE*

*Abstract*— **Driven by the special requirements of the Low-power and Lossy Networks (LLNs), the IPv6 Routing Protocol for LLNs (RPL) was standardized by the IETF some six years ago to tackle the routing issue in such networks. Since its introduction, however, numerous studies have pointed out that, in its current form, RPL suffers from issues that limit its efficiency and domain of applicability. Thus, several solutions have been proposed in the literature in an attempt to overcome these identified limitations. In this survey, we aim mainly to provide a comprehensive review of these research proposals assessing whether such proposals have succeeded in overcoming the standard reported limitations related to its core operations. Although some of RPL's weaknesses have been addressed successfully, the study found that the proposed solutions remain deficient in overcoming several others. Hence, the study investigates where such proposals still fall short, the challenges and pitfalls to avoid, thus would help researchers formulate a clear foundation for the development of further successful extensions in future allowing the protocol to be applied more widely.**

## I. INTRODUCTION

The ever-tighter integration of physical world with computing has given birth to a new communication paradigm referred to as the Internet of Things (IoT) [1][2]. One of the building blocks of the IoT is the *Low-power and Lossy Network* (LLN), a collection of interconnected embedded devices, such as sensor nodes, typically characterized by constraints on both node resources and underlying communication technologies [3][4]. Node constraints may include restrictions on power, processing and storage, while the communication system is subject to high packet loss, frame size limitations, low data rates, short communication ranges and dynamically changing network topologies [5][6][7]. Such limitations render the development of efficient routing solutions for LLNs difficult, a task made still more arduous by the potential large-scale deployments of such networks, anticipated to comprise thousands of nodes or more [8] [9].

As a major enabling component of IoT systems, LLNs have recently attracted much attention from industry, academia and standards bodies, with the goal of developing routing solutions that guarantee efficient use of limited network resources. In 2009, as a part of the Internet Engineering Task Force (IETF) efforts, it was concluded that conventional ad hoc routing protocols such as AODV [10] are too inefficient to satisfy LLNs' unique routing requirements [7]. Consequently, numerous routing solutions and primitives targeting LLNs were suggested, including e.g. CTP [11], and Hydro [12]. Ultimately these attempts led to the standardization by IETF of the IPv6 Routing Protocol for Low power and Lossy Networks (RPL) [13][14] in an effort to augment appropriately previous routing solutions. Since then, several studies have reported that RPL suffers from limitations that may harm its efficiency and a good deal of research has been directed at addressing them. However, there have been few comparative assessments of the effectiveness of such efforts.

### A. Contribution of the Survey

Our main contribution in this survey is to provide the research community with a solid piece of work that extensively surveys, discusses and analyzes research efforts made into addressing the RPL protocol limitations. This is in addition to providing an insight into the different routing requirements that based on them RPL protocol is specified. In nutshell, the contributions of this survey can be summarized as follows:

- We provide a thorough background on LLNs environments, communication technologies and different LLNs routing requirements within the context of IoT applications.
- We present a comprehensive overview of RPL's limitations and drawbacks reported in the literature that are related to its core operations (i.e. routing selection and optimization, routing maintenance operations and *downward* routing).
- We provide an extensive survey and an in-depth analysis of research efforts made to address the limitations of RPL and assess where they still fall short of.
- We present an outline of the future research directions and open issues.

The main findings of this survey is that RPL's extensions were only partially successful in addressing its limitations. Hence, a further research into overcoming RPL's limitations that are not fully addressed is needed. We articulate that the main issues that RPL's extensions have not addressed successfully are the efficient construction of *downward* routes, load-balancing and metric composition.

### B. Overview of Related Survey Articles

In the literature, there are a few surveys on RPL within the context of LLNs [15]-[22]. However, the majority of these surveys have focused on analyzing the limitations of the protocol itself rather than assessing various research efforts made towards overcoming such limitations, which is the main objective of this survey. In [15], the authors reviewed the relevant research efforts pertaining to the implementation, performance evaluation, and deployment of RPL. Nevertheless, their survey did not discuss the research efforts made toward overcoming RPL limitations. The study in [16] described the standard limitations without analyzing any of the recent studies proposed to mitigate them. Although the authors in [17] reviewed and analyzed some contributions that enhanced protocol operations, these are restricted to mobility extensions and the authors did not consider RPL's core operations aspects. The authors in [18] introduced a survey that presents the history of research efforts in RPL from 2012 to 2016. In particular, the authors investigated how RPL has been used and evaluated in the context of LLNs; however, their survey is different in two primary aspects, namely, their primary focus and the scope. The primary focus of their article was to report on the success of RPL itself as a routing protocol. In contrast, our article primary focus is to report on the success of the state-of-the-art solutions that were proposed in the literature to overcome RPL limitations. Yet, we have devoted an ineligible portion of our survey to report on the success of RPL (i.e. elaborating on RPL limitations and weaknesses), the scope and the depth of both surveys articles differ significantly. While our survey tackles the issues related to routing maintenance, routing selection, optimization mechanism, downward routing, under-specification of some RPL aspects, incompatibility between RPL modes, memory limitations among many others, with providing a classification for RPL limitations under those categories, their survey tackled the issues related to upward routing, multi-sink, interoperability, multicast, interference, load-balancing, downward routing, mobility and security. The authors in [19] overviewed RPL's key metrics, features and OFs. However, they surveyed only a few number of articles related to RPL's objective functions without providing an in-depth analysis for the feasibility of reported studies. In addition, the authors have never analyzed any of the state-of-art solutions proposed to mitigate RPL weaknesses in terms of routing maintenance, and downward routing, nor they elaborated on the problems themselves. The study in [20] has analyzed whether RPL has satisfied the original requirements defined by the IETF related to mobility, traffic patterns, resource heterogeneity, scalability, and reliability. In the light of this analysis, they highlighted the current IoT trends and new requirements that may challenge the future adoption of RPL. However, the study did not assess any of RPL's extensions proposed in the literature to overcome its challenges, which is the core goal of our survey.

Several other surveys in the wider field of routing protocols are existed [23]-[29]. One of the previous studies in this regards is the survey presented in [23], which surveyed around 13 inter-domain, and intra-domain multicast routing protocols. The authors also provided a comparative study among the surveyed protocols in terms of multiple features such as memory requirements and latency among several others [23]. A taxonomy of geocast routing protocols in ad-hoc networks was introduced in [24]. The authors classified these protocols (i.e. geocast routing protocols) based on network structure they support and the presence of flooding mechanism. They also conducted a comparison study among the reported protocols in terms of message/memory complexity, robustness and delivery capability in partial networks. However, apart from being devoted for ad-hoc networks, this survey is outdated. The authors in [25] provided a comprehensive survey of opportunistic routing protocols in WSNs. In particular, they discussed the limitations, features, variations and the building blocks of opportunistic routing protocols categorizing them into five classes, namely, optimization-based routing, probabilistic routing, geographic routing, link-state routing, and cross-layer routing. The authors also identified the key open research issues related to optimization, deployment and design of such protocols. However, the survey did not discussed the routing protocols targeting specifically the LLNs environments. A comprehensive survey of cooperative routing protocols in WSNs is presented in [26]. The authors introduced a taxonomy of such protocols in terms of centralization optimality, and objective function. They also conducted a performance evaluation of a representative set of cooperative routing protocols highlighting their key challenges. However, this survey also did not elaborate on routing protocols developed specifically for LLNs. The authors in [27] introduced a survey that reviews the multipath routing and provisioning protocols in wired networks. They presented a layer-based overview of such protocols highlighting their key benefits, challenges, and drawbacks and discussing related open research issues. In [28], the authors presented a new taxonomy for Multipath Routing Protocols in Real-Time Wireless Multimedia Sensor Networks (WMSNs) highlighting their advantages and disadvantages. They pointed out that multi-constraint multi-path routing in WMSNs is still an open issue that needs to be addressed by future research efforts. The authors in [29] conducted a comprehensive survey of QoS routing algorithms in SDN networks. A four-dimensional evaluation framework based on topology, scalability and delay was proposed in this survey to evaluate and compare the reviewed QoS routing algorithms. Nevertheless, these last surveys are not targeting the routing issue in the LLNs.

Compared with the surveys in literature, we aim at providing a deep understanding and thorough discussions on the core technical bottlenecks encountered in the design of RPL routing protocol and their corresponding solutions. Moreover, this survey will discuss whether the solutions have overcome the reported limitations and drawbacks. Uniquely therefore, this paper surveys key elements of the latest attempts to overcome the standard limitations of RPL pertaining to its *downward* traffic, objective functions and routing maintenance operations. We seek to answer the following question: have the research attempts to address the limitations of RPL succeeded, or do more need to be done?

### C. Organization of the Survey

The study is broken down into five parts. Firstly, we introduce LLNs and their associated unique challenges and requirements within the context of IoT applications in Section II. Section III introduces the technical details of the RPL protocol. We devoted Section IV to present a comprehensive overview of RPL's limitations and drawbacks related to its core operations reported in the literature. In Section V, a through discussion of the research efforts that aspire to extend the protocol and overcome its limitations is presented, analyzing their common pitfalls and challenges. The research directions and open issues not yet fully addressed are highlighted in Section VI. Finally, we conclude the paper in Section VII.

## II. LOW-POWER AND LOSSY NETWORKS

In this section, we provide a thorough background on the LLNs characteristics (Section II-A), communication technologies (Section II-B) highlighting the key challenges and routing requirements in such networks (Section II-C and D).

### A. LLN Characteristics

The term Low-power and Lossy Networks (LLNs) was introduced by the standardization bodies to refer to a class of wired and wireless networks where the hosts are tightly constrained in their resources as well as communication technologies [30]. While the resources limitations include restricted power reserves and restricted processing and storage capacities, the underlying communication technologies may exhibit low data rates, highly asymmetric link characteristics, high data loss and high variability of data loss (lossy links), and short communication ranges. A typical LLN may comprise of anything from a few routers to thousands of restricted-resource actuators and sensor motes with some routing capabilities connected to the external world (e.g. Internet) through a special *LLN Border Router* (LBR) that has no such restrictions itself [30]. The Architecture of a typical LLN is depicted in Figure 1. The LLN hosts generally exhibit similar characteristics; however, differences may exist in computing and storage capabilities of nodes. In this regard, IETF has classified sensor nodes, based on their capabilities, into three classes: 0, 1 and 2 [30]. Class 0 devices are severely constrained in terms of memory and processing with no more than 10 KiB of memory: they are incapable of carrying out communications without the help of a gateway node [30]. Class 1 devices are less constrained in terms of memory and processing capabilities, have the capacity to run a lightweight protocol stack and carry out communications with other hosts without requiring a gateway node. Finally, Class 2 devices are the least constrained in terms of memory or processing capabilities and have the capacity to support a protocol stack similar to that used in traditional computers. However, even Class 2 devices can gain benefit from running a lightweight stack since more application resources will be available if fewer resources are used for networking [30]. This also has benefits in reducing development cost and supporting the interoperability between the three classes [30].

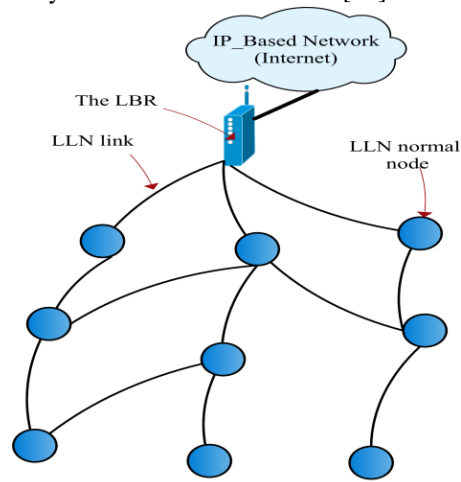

Figure 1. A Low-power and Lossy Network Architecture (LLN)

### B. LLN Standards and Radio Technologies

In order to facilitate the efficient deployment of LLNs in the context of IoT, several standards and radio technologies have been developed by different standardization bodies and research communities. In this section, we aim at providing an overview of these main standards and technologies. This includes the IEEE 802.15.4 and the IETF 6TiSCH standards, which address the issues related to the physical and MAC layers of the networking stack. It also includes an overview of the 6LoWPAN standard that provides an adaptation layer between the IEEE 802.15.4 standard and layer 3 protocols (i.e. IPv6 and RPL). In addition, an overview of other radio technologies within LLNs environments is also provided.

#### 1) IEEE 802.15.4 (Layers 1 and 2):

In order to satisfy the special requirements of the Low Rate Wireless Personal Area Networks (WPANs), an initial version of the IEEE 802.15.4 standard [31] was introduced in 2003 by IEEE 802.15 WPAN™ Task Group 4 (TG4). This version [31] defines the operations of two optional PHYs (layer 1) in different frequency bands with a very simple MAC layer (layer 2). The standard was then revised and amended several times specifically in 2006, 2007 and 2009. All these amendments were finally rolled out in a new version into a single document in 2011[32]. Pertaining to the technical features, the IEEE 802.15.4 has a maximum transmission rate of 250 Kb/s and a *maximum transmission unit* (MTU) of 127 bytes; however, only up to 116 bytes are available for an upper layer protocol [33]. At the MAC layer, the Carrier Sense Multiple Access with Collision Avoidance (CSMA/CA) scheme is used by to govern the access to the wireless medium [33]. Thanks to the efficiency of the standard, many of the recently specified upper layer networking stacks including 6LowPAN, ZigBee, and WirelessHART are built on the top of IEEE 802.15.4 [33].

### 2) 6LoWPAN (Layer 2.5)

Due to the restrictions imposed by the LLN devices and their underlying technologies, initially the Internet Protocol version 6 (IPv6) [34] was considered to be too resource intensive for such constrained devices [35]. Alternative proprietary solutions tended to implement complex application gateways to translate the non-IP format, understood by such networks, to the IP world [36]. However, various issues limited the adoption of gateways technology and led to a re-evaluation of the suitability of IPv6 for the LLNs [37]. In this new vision, the LLNs are no longer seen as isolated systems, i.e. proprietary solutions, rather they are seen as a key enabling technology for the ever-growing Internet of Things (IoT) paradigm where multitudes of identifiable smart objects, including smartphones, computers, laptops, actuators and sensors, are integrated into the Internet [38] [39]. However, the eventual LLN transition to the IPv6 world did not automatically resolve the old concerns about the demand on device resources and underlying communication technologies. For instance, while the key IEEE 802.15.4 medium access standard (layer 2) can only support a *maximum* transmission unit (MTU) of 127 bytes, the IPv6 protocol, which operates at layer 3, requires a *minimum* datagram size of 1280 bytes, approximately ten times greater [16]. In order to address such obstacles, the IETF commissioned the "IPv6 over Low Power Wireless Personal Area Network (LoWPAN) working group" to generate protocols that ensure smooth integration between LLNs and other networks running IPv6 [33]. These efforts culminated in specifying a new standard that allows for the IPv6 packets to be carried within the IEEE 802.15.4 MAC layer named, 6LoWPAN [4]. This has been enabled by identifying an adaptation layer, sometimes referred to as layer 2.5, between the IEEE 802.15.4 MAC layer (layer 2) and the IPv6 network layer (layer 3) as illustrated in Figure 2a. In particular, the 6LoWPAN adaptation layer defines the mechanisms of IPv6 header compression, IPv6 packet fragmentations and reassembly so that the IP datagram can be carried smoothly within the IEEE 802.15.4 frames.

### 3) IEEE 802.15.4e TSCH and IETF 6TiSCH (Layer 2)

Due to the single-channel related unpredictability of the IEEE 802.15.4 CSMA/CA in multi-hop networks, and to cope with the resource-constrained nature of LLNs, the IEEE introduced the Time-slotted Channel Hopping (TSCH) mode as an amendment to the MAC part of the IEEE802.15.4 standard in 2012 [40]. This new mode combines the TDMA (Time Division Multiple Access) with the channel hopping with the goal to improve both the energy efficiency and reliability [40][41]. While the TDMA scheduling minimizes the contention, and thus providing more efficient energy consumption, the channel hopping enhances the network reliability and mitigates the effect of channel fading [42][43]. In order to integrate the TSCH MAC protocol with IPv6 LLNs especially for industrial applications, the IETF chartered the "IPv6 over the TSCH mode of IEEE 802.15.4e" (6TiSCH) working group to enable IPv6 on the top of TSCH mode [43]. The 6TiSCH defines the 6TiSCH Operation Sublayer (6top) that specifies how nodes can communicate to add or delete cells and when such an addition and deletion can occur. The 6TiSCH also defines a set of distributed scheduling protocols that manage the allocation of resources. At the time of writing, the 6TiSCH is still active with two RFCs and five Internet-Drafts. These IEEE and the IETF joint standardization efforts have given birth to a modified 6LoWPAN stack named the 6TiSCH stack shown in Figure 2b.

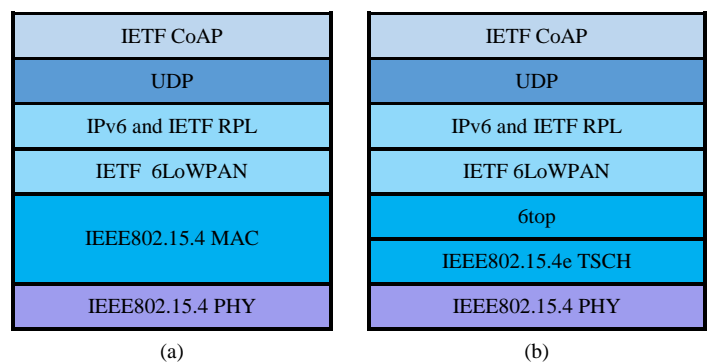

Figure 2. 6LoWPAN Stack (a) and 6TiSCH Stack (b)

### 4) Other Communication Technologies

**Bluetooth Low Energy (BLE) (layer:** BLE is WPAN technology designed for very low power operation, and is optimized for data transfer solution [44]. The recent specification of BLE (Bluetooth 5.0) [44] provides support for a data rate of up to 2 Mb/s within a short range (up to 200 meters) with multiple network topologies, including peer-to-peer, star, and mesh [45][46].

**Power-line Communication (PLC)[47] (Layer 1 and 2):** The IEEE 1901.2 specifies communications for low-frequency narrowband power line devices. It relies on re-using the existing electrical wires to provide communication capabilities supports a data transmission rate of up to 500 kb/s [47] [48]. Compared to its wireless counterparts, the PLC has the longest communication range, which is only limited by the length of the underline electrical cables [33].

**Wi-Fi HaLow (Layers 1 and 2):** To support the emerging concept of IoT networks, the IEEE 802.11ah Task Group introduced a new communication technology in 2016 named Wi-Fi HaLow based on the IEEE 802.11ah standard [49][50] [51]. The Wi-Fi HaLow supports data rates of at least 100 Kb/s with a communication range of around a kilometer [51].

### C.  Unique Routing Challenges in LLNs

The design of efficient routing protocols for LLNs is driven by the unique characteristics of such networks. The limited memory and processing resources, low data rates and limited power supply in the majority of devices along with the lossy nature of interconnects (links), all need to be addressed. In the following, we shed light on some of the routing process design issues arising in the context of LLNs.

**Diversity of Applications**: Several applications are envisaged to run under the umbrella of LLNs including home/building automation, industrial applications, environmental monitoring, military applications, etc.  These diverse applications exhibit characteristics [52] [53] and, consequently, different requirements in terms of power consumption, convergence time, traffic overhead, reliability, latency or other performance metrics. Hence, a big challenge for a LLN routing protocol is to accommodate all of these diverse and conflicting requirements within the application's resource budget [53].

**Communication Patterns:** The dominant communication pattern in LLN applications is the Multipoint-to-Point (MP2P) [14], in which data is gathered by a group of sensors and reported to a common destination called the LBR or the sink. Other communication patterns also exist, including the Point-to-MultiPoint (P2MP), where the sink sends data to the associated sensor nodes and the Point-to-Point (P2P) in which a sensor node communicates with one other in the network [54][55]. This diversity in communication patterns represents another challenge when designing LLNs routing protocols.

**Reporting Model:** The data communication models in LLNs vary widely, but are roughly classified into three categories, query-based, event-based and time-based [52] [53]. In the query-based model, data is only reported upon the receipt of an explicit query. In the time-based model, sensing devices report their data of interest periodically at pre-specified time. In the event-based model, sensing nodes only report their readings upon detecting abrupt and significant changes in the value of data of interest. Hybrid models combining two or more of these are also encountered [53]. The efficiency of routing protocols in terms of route stability and power consumption is highly sensitive to the reporting model.

**Scalability:** It is envisioned that LLNs will operate in deployments of different densities, ranging from a few neighbors per node to hundreds [54]-[59]. A protocol should thus be able to handle all cases within the viable range and its parameters should be dynamically tuned according to what it encounters in practice [53]. In other words, the scalability is a design issue that should be satisfied by an LLN protocol.

**Scarcity of Resources:** The resource-constrained nature of LLNs imposes a new set of restrictions on developing efficient routing protocols and primitives. Generally speaking, the small-battery capacity of a sensor node is the most restrictive factor and must be carefully considered [60].Thus, a routing protocol should opt to send just enough updates to ensure the freshness of the constructed routes while maintaining low-power profile. 'Just-enough' updates can vary from transmitting one update every second to a bulk transmission every few minutes, depending on the current conditions of the network and the type of application so as to ensure that the application energy budget is met [53].

**Links Unreliability:** LLNs are characterized by lossy and unreliable links, and an update is not guaranteed to reach its destination from its first transmission [61][62]. In some cases, the link loss rate in a network cannot be predicted beforehand and, even worse, the same link may exhibit different loss rates over time due to factors such as collisions at the receiver, the hidden terminal problem and interference with the radio transmitters of neighboring nodes [61]. However, there are still cases where an *a priori* loss rate can be roughly predicted depending, for example, on the statistics of previous deployments. Hence, a routing protocol should have the capacity to operate efficiently under such unreliable conditions.

**Mobility and Network Dynamics:** The sensor nodes in LLNs are conceived to be stationary in the vast majority of scenarios, however, there are still cases in which there are a considerable number of mobile nodes [54][55][56][57]. For instance, in health monitoring applications, the usual mode of deployment is mobile because sensor nodes are attached to the human body to monitor health conditions remotely while the subjects go about their business [64]. Therefore, general routing strategies must take account of possible node mobility.

### D.  LLNs Routing Requirements

The introduction of 6LoWPAN emphasized the need for additional IPv6-based LLNs routing solutions , thus, soon afterwards, IETF commissioned the *Routing over Low power and Lossy networks* (ROLL) working group to design such routing solutions [54]. Hence, the ROLL working group recognized that a wide range of application areas exists in LLNs each with its own routing requirements. Its first objective, therefore, was to define the routing requirements for four anticipated application areas, namely, Home Automation [54], Building Automation [55] Industrial LLNs [56], and Urban LLNs [57]. A discussion of these areas is now presented and a summary of these areas routing requirements is shown in Table 1.

**Home Automation:** Recently, the usage of sensing devices and actuators has increased in smart home applications. A modern home automation application will typically encompass both sensors, such as gas detectors, and actuators, such as heating valves [54][58]. These applications are designed to allow for the electrical devices at home to be connected to an IP-based system that controls these devices based on some input values from the end-users.

Table 1. LLNs Routing Requirements. Latency is the time taken to deliver a packet from the source to its intended destination; Convergence is the time taken for a node to re-establish end-to-end connectivity with other nodes in the network. Network-Scale is the number of the envisaged nodes in the network. Hops is the number of hops between the LBR and the farthest node. The mobility is the act of changing the locations of the sensor nodes or the LBR in the network field.

| Requirements | Home Automation | Building Automation | Industrial | Urban |
|---|---|---|---|---|
| Latency | Real-time, alarm and light control applications:<250 ms Other Applications: tens of seconds | Real-time, alarm and light control applications:<250 ms Other Applications: tens of seconds | Tens of milliseconds to seconds based on the type of the application | Variable based on the type of application |
| Convergence | Mobile: few seconds Fixed : less than 500 ms Subjected to: up to 250 nodes and four hops. | Fixed : less than 5 seconds Mobile: less than 10 seconds | Newly added device: within tens of seconds or several minutes Subjected to: tens of devices | Reporting Applications: lower than the smallest reporting interval |
| Network Scale | Typical: 10- 100 nodes Max: 250 nodes | Typical : 100- 1000 nodes Max: 10000 nodes Subnetworks: Up to 255 nodes | Typical: 10- 200 nodes | Max: $10^7$ Subnetworks: $10^2$- $10^4$ |
| Hops | Typical: 5 hops Max: 10 hops | Typical: 5 hops Max: 10 hops | Max: 20 hops | Several hops to several tens of hops |
| Mobility | Needs to be supported | Needs to be supported | Needs to be supported | Generally fixed locations |
| Traffic Pattern | P2P (prevalent), P2MP, MP2P | P2P (30%), MP2P and P2MP (70%) | MP2P (prevalent) P2MP (rare) P2P (rare) | MP2P (prevalent) P2MP and P2P (moderate) |
| Communication Model | Query-based (prevalent) Regular-based Event-based | Regular Query-based Event-based | Periodic , Query-based and Event-based | Regular (prevalent), Query-based (occasionally) Alarm-based(rare) |

Typical use-cases of home applications include: at-home health reporting and monitoring; lighting, central heating and air conditioning remote control; alarm systems (e.g. carbon-monoxide, smoke, fire detection, panic button, etc.) [54].

The majority of devices (sensors and actuators) in a home-operated network are stationary; however, there are scenarios where mobile devices are present, such as the wearable healthcare devices used to collect bio-medical signals remotely and also home applications controlled using a remote controller that moves from one location to another at random [54]. Supporting mobility is, therefore, a necessary requirement for the successful deployment of home automation networks.

The traffic patterns within this category vary widely [54] [58]. For example, the MP2P is used for communicating health conditions (e.g. blood pressure, temperature, insulin level, weight), while the P2MP model is appropriate for a lighting control system, where a central device sends control commands to associated devices [58]. However, the P2P traffic pattern has dominance here as most of the traffic in home-automation applications is generated by wall controllers and remote controls to their associated light or heat sources [54][58].

It is envisioned that a typical home automation network will be composed of tens of nodes with a maximum hop separation of a few nodes, and typically network diameter of five hops [54]. Many devices will be battery-powered so power consumption should be kept minimal to prolong network lifetime [54]. The majority of devices in home automation networks are likely to be Class 0 nodes (e.g., wall switches) with the rest of typically Class 1 [54]. The routing protocol for stationary devices has convergence requirements of no more than half a second, relaxed to four seconds at the presence of mobility [54]. For instance, a remote control appears unresponsive if it takes more than a second to pause the music [54][58].

**Building Automation:** These systems are deployed in a large

set of commercial buildings such as hospitals, colleges, universities, high schools, governmental and manufacturing facilities [55]. They typically enable automatic control of a commercial building's lighting, air conditioning, ventilation, fire-response and physical security among other systems [55]. The main purposes behind building automation are: reducing operating costs and energy consumption; enhancing occupant comfort; improving building service quality and the life cycle of utilities [55].

As with home automation, the majority of nodes in building-automation networks are Class 0 and Class 1 stationary devices with a small proportion of mobile nodes [55] [58].

It is expected that 30% of the traffic in building networks will be P2P with a typical frequency of one packet per minute [58]. For example, in a temperature-controlling application, a sensor will unicast periodically (e.g. each minute) of temperature readings to its associated controller and expects an acknowledgment unicasted from that controller. The MP2P and P2MP will account for 70% of traffic in this domain [55] [58]. This is because that most of the messages in building applications are directed toward an aggregation point and then routed off the LLN for further processing (MP2P). In addition, an acknowledgment is unicasted from the destination to the respective sender (P2MP).

The number of nodes in a building network is likely to be of larger size than in the domestic equivalent. However, a large building network would typically be divided into subnetworks of no more than 255 nodes to ensure that critical systems such as air conditioning and light systems are not vulnerable to global failures [58].

The latency requirements in building automation systems is somehow similar to the latency requirements in home-automation applications. However, many of the applications in this category are mission-critical (e.g., security fire) that are very sensitive to delay and require in-time delivery of messages [58]. Network devices (sensors and actuators) might be mains-powered, battery-less, or battery-powered [55].

**Industrial LLNs:** Industrial applications of LLNs enable plant and factory workers to manage remotely multiple control units at the site as well collect large amounts of information. Many application scenarios fall under this category, and they can be roughly classified into two different segments known as Factory Automation and Process Control [56]. Process Control applications target fluid products such as liquid chemicals, gas and oil, whereas Factory Automation is concerned with individual products such as cars, toys and screws [56].

All three communications patterns (P2MP, P2P and MP2P) will usually be present; however, the predominant traffic pattern is expected to be MP2P.

The majority of applications will comprise tens of field devices and forwarders with a few hop to reach a backbone router [56]. LLN devices in industrial networks may use a variety of sources to provide power: while some will be line-powered, the majority will be battery-operated with lifetime requirements of at least five years [56].

The issue of mobility in this category is more complicated than in the home and building scenarios and velocities of up to 35kph are possible [56]. For instance, some field devices may be located on moving objects such as cranes.

Since critical classes within this category are not expected to be handled by LLN routing protocols, the requirement of rapid convergence is somewhat relaxed [56]. It is stated in [56] that a routing protocol should converge within a few minutes of adding a new node with a latency of no more than ten seconds when delivering packets via established routes.

**Urban LLNs (U-LLNs):** These networks are dedicated to measuring and reporting a wide gamut of data in outdoor urban environments with the primary goal is to improve inhabitants' living conditions and monitoring compliance with environmental present; however, the predominant traffic pattern is likely to be MP2P [57]. Typical applications include the monitoring of meteorological conditions or pollution and allergen concentration in specific regions.

The dominant communication paradigm is the MP2P, as most of the traffic in this category will be generated by the sensor nodes and directed to the LBRs [57]. For example, the sensing nodes that gather temperature readings could communicate data every hour or every day. The P2MP model is also present: for instance, a query statement can be launched by a central unit to request pollution level readings from a group of sensors in a specific region [57].

Although most sensing devices in this category are expected to be stationary, the dynamicity of the network is not negligible, due to node disappearance, disassociation and association, in addition to perturbations of node interconnects [59].

Scalability represents the biggest concern in this category as the extensive measurement spaces in urban environments can result in very large networks. As currently imagined, an urban network will comprise more than a hundred nodes but sizes of tens of thousands, perhaps even millions of nodes, may be reached in the future [57] [59]. Although an urban network node cardinality is expected to be of the order of 5 to 10, examples of nodes with hundreds of neighbors may be encountered [57]. In addition, the physical distances between network devices can span from hundreds of meters to as much as a kilometer. Thus, it is unlikely that any field device will be able to reach its border router in a single hop and multi-hop distributions composed of as many as tens of hops may be unavoidable [57]. A mix of sparse and dense deployments in urban networks is expected: for instance, hundreds of devices may be presented in close proximity within one building in an urban area, whereas sparse node distributions, with low cardinality, would be the norm in sparsely built-up areas [59]. Devices may be powered using a variety of mechanisms, including: non-rechargeable batteries; rechargeable batteries with irregular or regular recharging; inductive/capacitive energy provision; or always-on (e.g. a powered electricity meter) [57]. It is likely, however, that the majority of nodes will use non-rechargeable batteries with lifetime requirements of 10-15 years [57].

Latency requirements in urban applications vary widely. For instance, for periodic traffic, latencies of up to a fraction of the reporting interval may be acceptable, while query-based applications will have somewhat more stringent requirements. Alert traffic is highly sensitive to delay and cannot tolerate a wait of more than a few seconds in the vast majority of cases [57].

## III. The IPv6 Routing Protocol for LLNs (RPL)

RPL is an IPv6-based proactive distance-vector routing protocol designed by the IETF community to fulfill the routing requirements of a wide gamut of LLN applications [14]. RPL is optimized particularly for data gathering purposes (i.e. MP2P traffic pattern), and it also provides a reasonable support for the P2MP traffic pattern, while providing an indirect support for the P2P pattern [14][15]. In this section, we provide an overview of the RPL's underline topology and its operations to build such a topology (Section IV-A). In addition, a discussion of RPL's standardized objective functions is given (Section IV-B) while the RPL routing maintenance mechanism is introduced in (Section IV-C). Finally, an overview of RPL's implementations in the literature is given (Section IV-D).

### A. RPL Topology and Operations

RPL organizes its physical network into a form of *Directed Acyclic Graphs* (DAGs) where each DAG is rooted at a single destination and is referred to as a *Destination-Oriented DAG* (DODAG) in RPL's terms [14]. The DODAG represents the final destination for the traffic within the network domain bridging the topology with other IPv6 domains such as the Internet [14][15]. It is referred to as the LLN Border Router (LBR) in the context of LLNs. RPL uses the term *upward* routes to refer to routes that carry the traffic from normal nodes to the LBR (i.e. MP2P) whereas routes that carry the traffic from the DODAG root to other nodes (i.e. P2MP) are called the *downward* routes. To build the *upward* routes, each node within the network must select one of its neighbors as its preferred parent (next-hop) towards the root. Similarly, each node willing to participate in the *downward* routing must announce itself to one of its parents, preferably the preferred parent. The details of building the *upward* and *downward* routes are given in the following subsections and an illustration of these operations are depicted in Figure 3. RPL uses the term *instance* to refer to multiple DODAGs that share the same routing policies and mechanisms. Multiple RPL instances may coexist concurrently in a specific physical topology and a node may join more than one instance at time. However, within each instance, a node is allowed to associate with only one root (DODAG) [14].

To exchange routing information needed to construct the network topology and routing paths, RPL introduces four ICMPv6-type control messages (excluding the security messages) as detailed below.

**DODAG Information Object (DIO):** the DIOs are used to carry the relevant information and configuration parameters that enable a node to discover RPL instance, join a specific DODAG, select a set of candidate parents, and maintain the DODAG [14].

**Destination Advertisement Object (DAO):** this control message allows a node to propagate its destination information upward along the DODAG to the DODAG root so that the downward routes from the DODAG root to its associated nodes can be constructed [14].

**DODAG Information Solicitation (DIS):** this message is used by an RPL node to solicit a DIO from neighboring nodes in order to join the DODAG [14].

**Destination Advertisement Object Acknowledgement (DAO-ACK):** the DAO-ACK may be unicast by a DAO recipient to the DAO sender to acknowledge the reception of that DAO [14].

### 1) RPL Upward Routes (Building the DODAG Topology)

The process of building the DODAG and *upward* routes is controlled by DIOs [13][14]. In addition to other routing information, the DIOs carry the *rank*, the relative position of an RPL node with respect to the DODAG root, and a routing policy called the *Objective Function (OF)* that specifies how an RPL node computes its *rank* and selects its preferred parent (the details of *rank* and *OF* are presented in Section III-B). Specifically, the construction of the DODAG is initiated by having the DODAG root multicasts DIO messages to its neighboring nodes announcing its *rank* and the OF that should be used [14]. This is depicted in Figure 3a. When receiving a DIO, an RPL node (a) adds the sender address to its candidate parents set, (b) calculates its own *rank*, (c) selects its preferred parent from the candidate parents, and finally, (d) updates the received DIO with its own *rank* and then multicasts the calculated rank to other neighboring nodes [14]. The node may also silently discard the received DIO based on some criteria defined in the RPL specification. This process lasts until all nodes have setup their routes in the *upward* direction towards the DODAG root as depicted in Figure 3b. The details of how RPL calculates the *ranks* of nodes is explained in Section III.B.

### 2) RPL Downward Routes

In order to facilitate P2MP and P2P communication patterns, *downward* routes must also be established and maintained. RPL uses the ICMPv6 DAOs messages for this purpose. An RPL node willing to announce itself as a reachable destination from the root point of view, unicasts a DAO to its preferred parent advertising its own destination prefix [13][14]. The processing of the received DAO by the parent relies on the current mode of operation advertised in the DIO messages. To this end, RPL has specified two modes for creating and maintaining *downward* routes, namely, *storing* (table-driven) and *non-storing* (source routing) [13][14]. In the storing mode, when a parent receives a DAO from one of its children, it: (a) stores the announced destination prefix locally in its routing table along with the DAO sender address, as the next hop to reach that destination; and (b) forwards the received DAO, in turn, to its own preferred parent to ensure the propagation of the advertised destination upward to the DODAG root [13][14]. For data-plane operations, classical hop-by-hop IPv6 routing is used.

In the non-storing mode, the advertised DAO carries also the address of the destination's parent in addition to the advertised destination prefix. Here, however, a parent receiving a DAO just forwards it to its own preferred parent without maintaining any routing state, until it is finally received by the DODAG root. Once the DODAG root receives the transmitted DAO, it maintains the received information in its routing table in the form of a child-parent relationships, used later by the data-plane to construct a source route for the intended destination [13][14]. Hence, when the root needs to communicate with a specific destination, it attaches the source route of that destination to the packet header and forwards the packet to the next hop. A forwarding node receiving that packet will simply inspect the source routing header to determine on which interface it should send the packet next [14].

RPL also provides a support for the P2P traffic pattern in which a node communicates with another node in the network. Hence, when a node needs to send a packet to another node within the DODAG, the packet is first forwarded upward the DODAG until it arrives at an ancestor that has a known path to the destination node. Then, the packet is forwarded downward the DODAG by that ancestor via the intermediate routers and finally to the destination node. A high-level illustration of these operations discussed above is depicted in Figure 3c while Figure 3d depicts the final DODAG of an LLN.

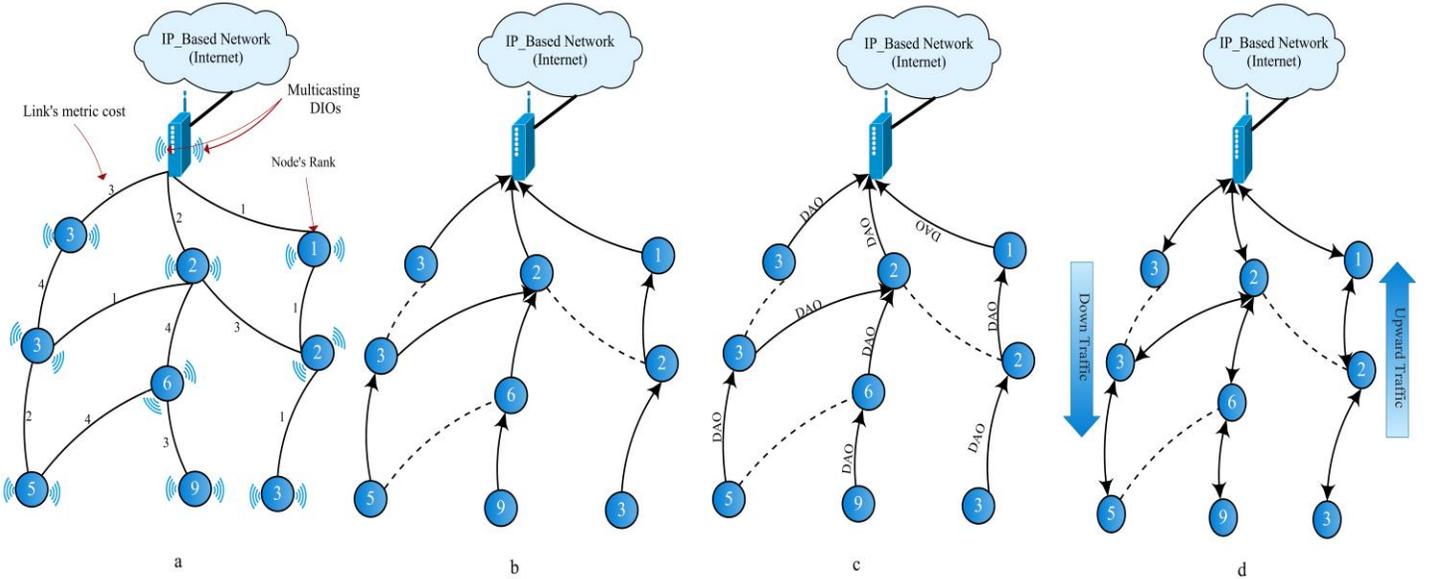

Figure 3. The operations of building the *upward* and *downward* routes: a) Propagation of DIOs and *rank* calculation initiated by the LBR (root), b) The DODAG topology and *upward* routes: where an arrow goes from a child to its preferred parent while a dotted arrow represents another candidate parent, c) The propagation of DAOs over already built DODAG to build the *downward* routes, d) The final DODAG with bi-directional communication capabilities (*upward* and *downward*)

## B. Objective Functions (OFs)

In order to meet the conflicting requirements of different LLN applications, RPL decouples the route selection and optimization mechanisms from the core protocol operations such as packet processing and forwarding [14]. Hence, the core of the protocol is centered on the intersection of these requirements, whereas additional modules are designed to address application-specific objectives such as minimizing the energy consumption or maximizing the reliability [65][66]. The term Objective Function (OF) is used to describe the set of rules and policies that governs the process of route selection and optimization in a way that meets the different requirements of various applications. In technical terms, the OF is used for two primary goals; first, it specifies how the *rank* can be derived from one or a set of routing metrics [67] (e.g. energy, hop count, latency, throughput, link reliability and link color[2]), second, it defines how the *rank* should be used for selecting the preferred parent [65] [66]. Currently, two OFs have been standardized for RPL, namely, the Objective Function Zero (OF0) [65] and the Minimum Rank with Hysteresis Objective Function (MRHOF) [67].

### 1) The Objective Function Zero (OF0)

The OF0 is designed to select the nearest node to the DODAG root as the preferred parent with no attempt to perform load balancing [65]. The *rank* of a node ($R_n$) is calculated by adding a strictly positive scalar value (*rank_increase*) to the *rank* of a

selected preferred parent ($R_p$) according to Eq. 1 and Eq. 2 as follows:

$$R_n = R_p + rank\_increase \tag{1}$$
$$rank\_increase = (R_f * S_p + S_r) * MinHopRankIncrease \tag{2}$$

where the *step-of-rank* ($S_p$) represents a value related to the parent link metric and properties such as the hop-count or the Expected Transmission Count (ETX), while the *rank* factor ($R_f$) and *stretch_of_rank* ($S_r$) are normalization factors [65]. The OF0 does not specify which metric/metrics should be involved in the calculation of rank increase. For parent selection, a node running OF0 considers always the parent with least possible rank as its preferred parent. OF0 considers also selecting another parent as a backup in case the connectivity with its preferred parent is lost [65].

### 2) Minimum Rank with Hysteresis Objective Function (MRHOF)

The MRHOF [66] is designed with the goal to prevent excessive churn in the network topology (i.e. frequent change of the preferred parent). In the MRHOF, a node calculates the path cost through each neighbor by adding up two components; the value of the candidate neighbor node's or link's metric and the value of the selected metric advertised in the Metric Container. After calculating the path costs of all candidate parents, a node selects the parent with lowest path cost as its preferred parent.

---

[2] The link color is an administrative 10-bit field Constraint (can be used also as a recorded metric) [67]. The meaning of each bit is to be defined by the implementer. For example, an implementer may set the first bit of the field to 1 indicating that this link is an encrypted link. Hence, when used as recorded metric,

the number of encrypted links might be reported along different alternative paths and a routing policy may be setup to prefer a path having the maximum number of encrypted links [67].

However, unlike OF0, MRHOF switches to a new parent only if the new minimum calculated path cost is smaller than the preferred parent's path cost by at least PARENT_SWITCH_THRESHOLD, which is the hysteresis part of MRHOF [68]. If multiple candidate parents share the same path cost, other tie-breaking criteria might be used [66] [68].

### C. Routing Maintenance (Trickle Timer)

One of the key design principles of the RPL is minimizing the routing control overhead and signaling cost in order to reduce energy consumption and enhance reliability. In this regard, RPL employs the Trickle algorithm [53] [69] to govern the transmission of the signaling traffic used to construct and maintain the DODAG. The basic idea behind Trickle is to adjust the frequency of message transmission based on network conditions. Trickle relies on two simple mechanisms to disseminate routing information efficiently. The first is to change adaptively the signaling rate according to conditions currently present in the network [53]. Specifically, Trickle increases the transmission rate when a change in routing information is discovered (i.e. an inconsistency is detected) as a mean to populate the network rapidly with up-to-date information [53] [69]. As the network approaches its steady phase, Trickle exponentially reduces the transmission rate to limit the number of transmissions when there is no update to propagate. The second mechanism used by Trickle is the suppression mechanism in which a node suppresses the transmission of its control packet if it detects that enough of its neighbors have transmitted the same piece of information, thus limiting redundant transmissions [53] [69].

### D. RPL's Self-healing Mechanisms

To protect against failures in the network, RPL provides self-healing mechanisms to detect and avoid loops and to repair the DODAG as follows:

#### 1) Loop Avoidance and Detection

In the distance-vector protocols including RPL, loops are a common problem that may form due to several reasons (e.g. loss of control packets) affecting negatively the performance of the network [14][15]. The rank-based routing used by RPL serves as a mechanism to avoid loops whereas the detection of loops is performed using another simple mechanism named Data-Path validation [14]. In the data-path validation, RPL injects some routing information in the transmitted data packets that indicates the direction of the flow (i.e. upward or downward) and the *rank* of the sender. Hence, an inconsistency between the direction of the packet and the *rank* relationship between the sender and receiver nodes is an indication of a possible loop [14]. For instance, if a node receives a packet moving downward from a higher-ranked node, then the receiving node can deduce that inconsistency has occurred, as a packet received from a higher-ranked node must only progress in the upward direction.

#### 2) DODAG Repair

For overcoming failures that may occur in the network such as loops, RPL uses two dynamic repair mechanisms named Global Repair and Local Repair [14][15]. In the Local Repair, a non-root node that detects an inconsistency (e.g. loop or link failure) should detach itself from the DODAG by announcing a rank of INFINITE RANK to poison its routes and then reattach to the DODAG as a new joining node [14]. In the Global Repair, totally a fresh DODAG topology is constructed. The global repair can be only triggered by the DODAG root upon detecting a failure in the network and it is instituted by incrementing the DODAG Version Number field within the DIO message [14].

### E. RPL's Security Features

As reported in [14] RPL usually relies on the underline link-layer mechanisms to support the security features of authenticity, integrity and confidentiality. However, at the absence of such mechanisms, RPL uses its own specified security mechanisms with three optional security modes have been specified by the standard as follows:

#### 1) The Unsecure Mode

In this mode, RPL control messages are transmitted without including any additional security features [14]. In this case, RPL relies on other layers security primitives to satisfy the security requirements of the network [14].

#### 2) The Pre-installed Security Mode

In the pre-installed mode, nodes are provided with pre-installed keys with which RPL secured messages can be generated and processed [14].

#### 3) The Authenticated Security Mode

Like the pre-installed mode, nodes are provided with pre-installed keys; however, they may only be used to join the instance as a leaf. A router joining a RPL instance will need to require another key from an authentication authority [14].

### F. RPL's Implementations

Several vendor and open-source RPL's implementations exist in the literature [70]-[81], however and as reported in [77], there is no such an implementations that implemented the full list of RPL specifications.

#### 1) RPL Open-Source Implementations

##### a) ContikiRPL

Contiki [71] [79] is a lightweight and open-source operating system designed specifically for the low-power resource-constrained IoT devices. Contiki features a highly optimized networking stack including several IoT standards such as 6LoWPAN and IPv6. It also features an implementation for the RPL standard fundamental mechanisms within a library called ContikiRPL. Both the OF0 and the MRHOF are implemented within the library with the OF0 uses the hop-count as its routing metric and the MRHOF uses the ETX. In addition, the latest version of ContikiRPL includes both the storing and the non-storing modes of RPL. In 2017, the authors of Contiki started a new fork of the Contiki operating system named Contiki-NG [78], which features two different implementation of RPL: RPL-classic, and RPL-light. RPL-classic has a code size of 227 KB whereas RPL-light has a relatively smaller code footprint of 204 KB. The main difference between the two implementations is that RPL-light do not implement some features that seems unnecessary according to the analysis of RPL in [77] such as the storing mode and the existing of multiple instances (e.g. only one instance has been supported that uses the MRHOF and ETX metric). However, all Contiki-based implementations of RPL do not include any of its security features.

### b) TinyRPL

TinyOS has its own implementation of the RPL standard named TinyRPL, which is deigned to be used with BLIP (the Berkeley Low-power IPv6 Stack). The last implementation of TinyRPL supports both the storing and non-storing modes of RPL with the default *upward* routes. It also supports the two standardized OFs (i.e. OF0 and MRHOF). However, TinyRPL supports only a single instance with multiple DODAGs whereas it lacks in any support for RPL security features. The codes size of Tiny RPL is smaller than that of ContikiRPL with only 113 KB.

### c) RIOT-RPL

RIOT [80], an operating system for memory-constrained low-power wireless Internet of Things (IoT) devices, has also its own implementation of the RPL standard named RIOTRPL [73]. RIOTRPL supports the two *downward* RPL's modes of operations; however, it only implements the OF0 with hop-count routing metric. It has a code size of more than 105 KB; however, it does not provide any support for the security modes of RPL [77].

### d) Unstrung

Unstrung is a user-space Linux-based implementation of the RPL protocol intended for wired/Ethernet backhaul networks and gateway systems [74][75]. It can run on laptops, multipurpose IoT nodes, access points and diagnostic devices [74]. The implementation is mostly written in C++ with a code size of 1 MB. While Unstrung supports the storing mode of RPL, it does not provide support for the non-storing mode.

### e) SimpleRPL

SimpleRPL is another user-space implementation of RPL for Linux-bases systems. It is written in python and has a code size of 228. Pertaining to *downward* routing, SimpleRPL supports only the storing mode without multicast [76][77]. In addition, only the OF0 with hop-count metric is supported by SimpleRPL with the capability to form only one DODAG. Like other implementations, SimpleRPL does not provide any support for the security features of RPL as it is expected to be run on a secure environment [76].

### 2) RPL Vendor Implementations

According to [77], several vendors have implemented their own versions of RPL including Samsung, Huawei, and Cisco. However, the available information about these implementations is very scarce as they are confidential. Only Cisco has revealed some of the implemented features in a form of configuration guide available online in [81]. Several features have not been implemented by Cisco including the secure mode of RPL and the non-storing mode. In order to cover a wide spectrum of uses in smart cities, Cisco implementation of RPL includes support for three OFs, namely, OF0, OF1 (latency) and OF1 (ETX) [81].

## IV. RPL LIMITATIONS AND DRAWBACKS

As the de-facto standard for routing in IoT networks, a plethora of recent studies have evaluated RPL performance reporting several limitations and pitfalls that need to be addressed [82]-[123]. In the next subsections, we elaborate on the key weaknesses and limitations reported in the literature related to RPL's OFs (Section V-A), RPL' *downward* routing (Section V-B) and RPL's routing maintenance (Section V-C). A summary of these limitations is presented in Table 2.

### A. Objective Function Limitations

In the section, the issues related to RPL OFs are discussed including the single-path routing, the under-specification of metric composition, and the implicit hop-count impact.

### 1) Single-path Routing

In RPL, once a preferred parent has been selected, all traffic will be forwarded through this preferred parent, as long as it is reachable, without any attempt to perform load balancing among other available parental candidates [13][14][16]. This behavior may drain the power of overloaded parents leading to network disconnections and unreliability problems, as it is likely that overloaded nodes will die earlier [63][103][104].

### 2) Under-specification of Metrics Composition

RPL supports the use of multiple metrics for routing with the possibility of optimizing the routes based on combining several metrics, however, no guidelines are provided on how such combination should be achieved [89]. Hence, relying on a single routing metric in the OF may satisfy one application requirement, yet also violate another [90][92][98]. For example, while the ETX routing metric allows the protocol to select the most reliable path, it may also result in early network partitioning due to the absence of a load-balancing mechanism that might protect vulnerable nodes from exhausting their battery power. The problem of unbalanced traffic is exaggerated by the fact that standard RPL permits forwarding the traffic through the preferred parent only, even in the case when several candidate parents are available to do the job [89].

Table 2. The summary of major RPL's limitations

| The problem | The module | Brief description | Side effects/ pitfalls |
|---|---|---|---|
| Incompatible modes for *downward* routing (Section IV-B-1) | Downward routing (Section IV-B) | The *downward* MOPs are not specified to understand each other. | Forwarding failure and network partitions |
| Memory limitations (Section IV-B-2) | Downward routing Storing mode (Section IV-B) | Each node must maintain the routing entries of all nodes in its sub-DODAG which might not be possible for memory-constrained nodes | Memory overflow jeopardizing reliability and scalability |
| Long source headers in the non-storing mode (Section IV-B-3) | Downward routing Non-storing mode (Section IV-B) | Transmitted packet must carry the addresses of all nodes to destinations | Higher overhead jeopardizing reliability and scalability |
| Under specification of DAOs emission (Section IV-B-4) | Downward routing (Section IV-B) | when a node should transmit its DAOs is unspecified | May lead to inefficient implementations |
| Listen-only timer (Section IV-C-1) | Routing maintenance timer (Section IV-C) | A node must wait for the half of the interval before transmitting a routing update | Slow convergence and load-balancing problems |
| Suppression mechanism Inefficiency (Section IV-C-2) | Routing maintenance timer (Section IV-C) | Node must suppress a specific routing update should it hear that a certain number of the neighbors have transmitted the same routing update | If not configured correctly, forming sub-optimal routes |
| Single-path routing (Section IV-A-1) | Objective Function (Section IV-A) | A node keep forwarding traffic to its preferred parent with no attempt of load balancing | No load balancing affecting negatively both reliability and energy efficiency. |
| Under-specification of metrics composition (Section IV-A-2) | Objective Function (Section IV-A) | No guidelines are specified on how to combine several metrics | Jeopardizing the capacity of the protocol to get the benefit of combining several metrics |
| Implicit hop-count impact (Section IV-A-3) | Objective Function (Section IV-A) | A path with better global quality (usually due to its less number of hops) may contains one or more links with critically low-quality links that undermine its apparent quality | May impact negatively any performance aspect |

### 3) Implicit Hop-count Impact

In RPL's objective function, the routing cost of a specific path is calculated by adding up the cost of its constituent links. Hence, a path with a large number of hops will appear more costly than another path with a relatively small number of hops even though the first path's constituent links might be of better quality [94]. This might be misleading when taking routing decisions as the path with the small number of hops would have a higher probability of being selected even though it might have one or more very low-quality individual links [94].

### B. RPL Downward Routes

According to the specification of RPL standard in [14], it is expected that MP2P traffic pattern will be the dominant pattern in the context of LLNs while other traffic patterns (i.e. P2MP and P2P) are expected to be less common. Adhering to these expectations, RPL optimizes it routes for the upward traffic in way that requires less overhead and minimized routing state. However, this has been achieved at the cost of somewhat inefficient construction of downward routes in terms of control overhead, routing state and path stretch [16] [87][113][114] [115] [116], resulting in some issues as follows.

### 1) Incompatible Modes for Downward Routing

Although RPL supports two different modes for *downward* traffic (i.e. storing and non-storing), the standard specifies that RPL-compliant deployments should use either the non-storing mode or the storing mode within the same instance [14] [114]. Hence, when nodes belonging to different instances running different modes of operation meet in the same RPL network, RPL permits nodes from one instance to join the other instance only as a leaf node, which gives the rise for several interoperability problems. For instance, consider the case when a node from one instance located in the middle of a forwarding path joins another incompatible instance as a leaf while it represents the only available next-hop to the DODAG root [114]. Hence, nodes downstream of the new node cannot now communicate with the root through it, since the leaf is not allowed to operate as a router and the network is thus partitioned in both the *upward* and *downward* directions [114]. One solution is to relax the restriction and allow nodes with different modes of operation to join incompatible instances as routers [114]. However, a forwarding failure may still occur in downward traffic as a router operating in storing mode will have no capacity to understand the source header of a packet sent by a non-storing peer [114].

### 2) Memory Limitations in the Storing Mode

RPL requires that every node running the storing mode of operation must maintain the routing state of all nodes in its sub-DODAG (a node's sub-DODAG represents the set of other nodes whose routing paths to the root are passing via that node) [14][114]. Although RPL is designed specifically for small and constrained-memory sensor nodes, the protocol has the ambition to handle dense networks comprising up to thousands of nodes. In such high-density networks, it is highly likely that the routing state need to be maintained will overflow the storage capacity of such constrained devices [116]. Hence, an overflowed node will be unable to accommodate all routing entries required to be maintained in its routing table, rendering several destinations in its sub-DODAG unreachable from the root point of view enforcing the root to drop the packets destined to such unreachable destinations [114] [116].

### 3) Long Source Headers in the Non-storing Mode

In the non-storing mode of RPL, the root is required to attach a source route header for each transmitted datagram in the *downward* direction [14]. However, RPL is designed to operate on link layers with a Maximum Transmission Unit (MTU) of 127 bytes [16][88]. Out of the 127 bytes available for the physical layer frame, a maximum of 46 bytes are reserved for the L2 header, a minimum of 2 bytes for the compressed IPv6 fixed header, and a fixed header size of 8 bytes for the attached source route. Considering this, only 71 bytes remain for the L3 datagram payload. Thus, a maximum of four hops in the source route header are possible as each IPv6 address has a fixed length of 16 bytes without compression. The compression techniques spelled out in [5][6] can allow up to 70 hops in the source header; however, as LLNs require IPv6 auto-configuration, a maximum of 8 bytes can be taken out of any compressed IPv6 allowing for a path length of maximum of eight hops from the source to the destination. This imposes a tight constraint on multi-hop transmission.

### 4) Under Specification of DAO Emission

A key issue in constructing RPL *downward* routes is that the timing of DAO transmission is not explicitly specified. This under-specification of DAO timing may lead to conflict and inefficient implementations of the protocol, consequently harming its performance [88]. For instance, the study in [83] has opted to transmit DAO messages periodically every 5 seconds, significantly increasing the control overhead compared to the ContikiRPL implementation [71], which transmits DAO messages based on the Trickle timers of DIOs. A conservative timing approach may lead to DAOs not being transmitted before old routes expire, affecting negatively the data-plane reliability [16]. Hence, an implementation that does not guarantee receipt of all DAOs from intermediate routers along a path would render the root unable to calculate the source route for that destination [16]. This is because the accurate calculation of a source route relies on all route segments advertised in the DAOs of its ancestors, up to the DAODAG root. Here, the root would again have no option but to drop all packets for the affected unreachable destination [16].

### C. Routing Maintenance (Trickle Timer) Limitations

As discussed above, the RPL standard specifies that Trickle must be used for routing information exchange and maintenance. The relying on Trickle has given rise to some issues as presented next.

### 1) Listen-only Period

A key issue in Trickle is the introduction of listen-only period in the first half of each Trickle's interval (*I*) [119][123]. The goal behind the listen period is to solve the so-called *short-listen* problem in asynchronous networks [53]. In a network with no listen-only period, a node may start sending its current DIO very soon after starting a new interval, a behavior that may result in turning down the suppression mechanism in the current and subsequent intervals, leading to significant redundant transmissions and limiting the algorithm scalability [53]. However, the listen-only period comes with its own shortcomings. Firstly, the period imposes a delay of at least *I/2 before* trying to propagate the new information. In an *m-hop* network, an inherited delay will progressively accumulate at each hop resulting in an overall delay proportional to the number of hops [119] [123]. Secondly, the listen-only period may result in uneven load distribution among network nodes with some nodes transmitting less than others do during the operational time [118]. In the worst-case scenario, the transmitting period of a node may substantially overlap with the listen-only period of a neighboring node, preventing the former from sending for a long time. A key issue here is that the blocked node may be one whose transmission is vital for resolving network inconsistences [119]. Furthermore, the absence of load balancing among Trickle nodes may render some routes undiscoverable even though they might be more efficient than those already active in the network [118].

### 2) Suppression Mechanism Inefficiency

Another issue with Trickle algorithm is related to its suppression mechanism. In order to lessen the control overhead in the network, Trickle suppresses the transmissions of control messages that seems to be redundant. It does so by counting the number of consistent messages that are received within a specific window and, then, when such a number surpasses a pre-configured redundancy constant *(k)*, it suppresses any further propagation of such received messages. However, studies have reported that the optimal setting of the redundancy constant is not a trivial task and relies greatly on the application scenario, in addition to that some issues may emerge if configured incorrectly [118][120]. For instance, it was shown in [118] that, if the redundancy constant is not configured correctly, the suppression mechanism might result in sub-optimal routes, especially in heterogeneous topologies with regions of different densities. This is attributed to the fact that Trickle is originally designed to disseminate code updates, which are quite similar in the context of reprogramming protocols. However, this is not the case in the context of routing as two routing update messages originated from different sources may carry different routing information and thus "suppressing one transmission or another is not always equivalent" [118].

## V. RPL'S ENHANCEMENTS: PROSPECTS AND PITFALLS

In this section, we survey the RPL's enhancements and extensions since its introduction related to its OFs, downward routing and routing maintenance. We provide an in-depth analysis of such extensions highlighting their key weaknesses. In particular, we survey the extensions of RPL's OFs (Section VI-A), the extensions of RPL's downward routing (Section VI-B) and the extensions targeting RPL's routing maintenance (Section VI-C). A classification of various enhancements is illustrated in Figure 4. For quick reference, Table 3 shows RPL's OF extensions and their weaknesses, while Table 4 shows RPL's downward routing and routing maintenance extensions with their key pitfalls.

## A. Objective Function Enhancements

Several efforts have been made to fill in the gaps presented in RPL's objective functions [89]-[112]. Most of these efforts focused on designing OFs with a composite routing metric to fulfill conflicting routing requirements in the same application domain. Introducing multipath routing as a mean of enhancing the efficiency of OFs is the focus of another class of studies.

### 1) OF Enhancements Based on Metric Composition

Several research studies have been done into overcoming the problem of the under-specification of metrics composition of the RPL standard. Hence, multiple mechanisms are proposed to combine the respected metrics including lexical, additive, hybrid, and fuzzy based composition. In the lexical composition, the selection of the parent is done based on the first composition metric and if two parents have equal values for the first composition metric, the second composition metric is used to break the tie [89]. In the additive composition, the weighted values of participating metrics are added to produce one composite value, which the selection of the preferred parent is done based on [89]. In the hybrid composition, both the lexical and the additive techniques are used to combine two or more metrics. The fuzzy based-composition is based on the concepts and principles of fuzzy logic. In the following, we discuss these extensions.

#### a) Hybrid Composition Enhancements

The authors in [89] propose lexical and additive composition techniques that combine two routing metrics to optimize multiple performance aspects. They pointed out that the monotonicity property of the combined metric must hold to ensure a loop-free routing protocol. When using an additive composition, the two component metrics must hold the same order relation to ensure validity of the composite metric. However, this restriction is not necessary when using lexical composition. The work proposes a combined Hop Count (HC) and Packet Forwarding Indication (PFI) metric, to construct shorter paths that avoid nodes acting maliciously or selfishly. Simulation results have shown that lexical combination of these two metrics gives better detection of misbehaving nodes and selection of reliable paths while showing comparable latency in comparison with the hop count metric only. The authors also show that combining Residual Energy (RE) and hop count metrics either in an additive or lexical manner results in better energy load distribution among nodes in comparison with hop count only.

The Scalable Context-Aware Objective Function (SCAOF) for agriculture low power and lossy networks (A-LLNs) is proposed in [93]. SCAOF combines the metrics of remaining energy, ETX, availability information, and hardware robustness (number of restarts) and affordable workload (the tendency of node to consume energy), in a way that guarantees the selection of a reliable path while avoiding nodes that have depleted their power reserves. This study also introduces the notion of ETX_Threshold and RE_Threshold in order to allow for a configuration that is consistent with specific application [93]. The proposed objective function is evaluated by means of simulations and testbed experiments and compared to RPL-ETX (the exact used OF is unclear) in terms of packet loss rate, routing table size, Round-Trip Time (RTT), overheads, path hop distance, packet delays, network churn, and network lifetime. It is shown that the developed protocol can reduce network churn, prolong network lifetime and enhance the quality of service of A-LLNs applications.

#### b) Additive Based Composition Enhancements

The study in [90] addresses the issue of RPL relying only on a single metric: energy or reliability. The authors in this study highlighted the problems of unbalanced traffic and the consequently uneven energy consumption distribution among network nodes in RPL. In particular, the study pointed that using

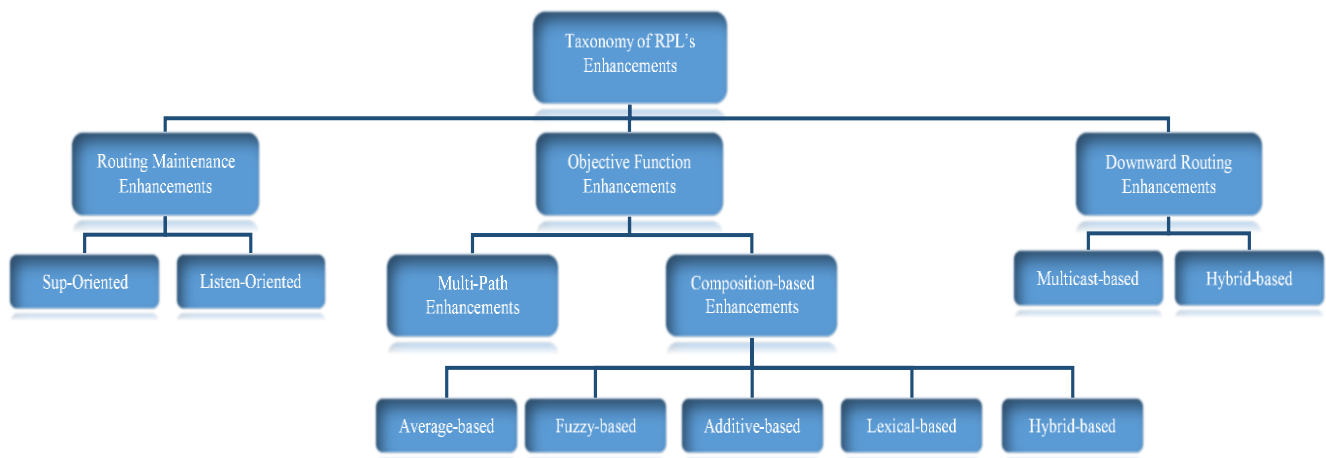

Figure 4. RPL's Enhancements Classification

ETX as single metric in RPL network would result in excessive use of some paths, especially those with high delivery rates. This excessive use of good-quality paths will result eventually in network partition and reduce the overall lifetime of the network [90]. If energy is selected as the sole routing metric, on the other hand, the reliability of the path might be impacted negatively. To balance energy consumption of nodes while providing highly reliable paths, the study proposes a weighted energy-oriented composite metric that takes into consideration a node's residual energy in addition to ETX. The study results show that energy consumption is balanced to some extent by the proposed technique, which enhances network lifetime by up to 12%.

An Energy Efficient and Reliable Composite Metric for RPL Networks is proposed in [92]. This composite metric takes into consideration both the reliability, represented by the ETX metric, and energy efficiency to balance energy consumption among nodes and enhance the network lifetime. The proposed metric is called the Lifetime and Latency Aggregateable Metric (L2AM). In particular, a node running L2AM, first combines transmission power of the link and a node's residual energy using an exponential function to produce what is called the *primary* metric. The ETX metric is then multiplied by the primary metric to get the composite metric overall cost: this is what must be minimized when selecting the preferred parent. For evaluation purposes, the proposed metric is compared to ETX RPL in terms of network lifetime and remaining energy. The results have shown that the L2AM outperforms ETX RPL by up to 56% in terms of network lifetime.

Table 3: The RPL's OFs extensions and their weaknesses. The metrics used in the table are HC (Hop Count), ETX (Expected Transmission Count), RE (Residual Energy), PFI (Packet Forwarding Indication), SI (Stability Index), ARSSI(Average Received Signal Strength Indicator), SPRR (Smoothed Packet Reception Ratio), SRNP (Smoothed Required Number of Packet retransmissions), PD (Propagation Delay), NC (Node Congestion), LC (Link Congestion), BDI (Battery Discharge Index) and RER (Residual Energy Ratio)

| Ref. | Metrics | Multi path | Type of Metric composition | Brief description | Limitation Addressed | Drawbacks | Minimum DIO Size increase (in bytes) |
|---|---|---|---|---|---|---|---|
| [89] | HC and PFI or HC and RE | NO | Lexical and additive | Combines HC and PFI for better detection of malicious nodes. Also combines HC and RE for load-balancing | Under-specification of metrics composition (Section IV-A-2) | No real testbed experiments Very low-quality paths still can be selected | +13 or +14 |
| [90] | RE and ETX | NO | additive | Combines RE and ETX for load-balancing | Under-specification of metrics composition (Section IV-A-2) | Only up to 6 nodes for evaluation. Very low-quality paths still can be selected | +14 |
| [91] | RE and ETX | NO | Lexical | Combines RE and ETX for building reliable and energy-efficient topology simultaneously. | Under-specification of metrics composition (Section IV-A-2) | No real testbed experiments Very low-quality paths still can be selected. | +14 |
| [92] | Transmit power, Energy and ETX | NO | additive | Combines RE and ETX for enhancing reliability and energy-efficiency with a mechanism to lessen the impact of highly depleted nodes. | Under-specification of metrics composition (Section IV-A-2) | Claimed reliability not reported nor justified No clarification on how DIO intervals selected. | +19 |
| [93] | RE, ETX , Link color and other context-aware metrics | NO | Lexical and additive | Combines RE, ETX, link color and other metrics to boost reliability while avoiding nodes that have depleted their energy. | Under-specification of metrics composition (Section IV-A-2) | Higher risk of fragmentation. Only up to 11 nodes for evaluation. Very low-quality paths still can be selected. | +21 |
| [94] | HC and ETX | NO | average | Combine the hop count and the ETX by taking the average of ETX to avoid long single-hop problem. | Implicit hop-count impact (Section IV-A-3) | The monotonicity property is not satisfied. Suffer from excessive churn. | +14 |
| [95] | HC, Number of children and distance to parent | NO | additive | Combine the distance, number of children nodes and the HC. | Implicit hop-count impact (Section IV-A-3) | High risk of fragmentation. No indication of the used simulation tool. | +20 |
| [96] | SI and ETX | NO | additive | Introducing new stability metric and combines it with ETX to build more stable and reliable topology. | Under-specification of metrics composition (Section IV-A-2) | Less control messages not only indicate stability, it may also indicates unreliable links. Unclear how the SI and EXT are combined. | +13 |
| [98] | HC, energy, ETX and delay | NO | Fuzzy-based | Combines hop count, energy, link ETX and delay to satisfy the most important requirements. | Under-specification of metrics composition (Section IV-A-2) | Higher risk of fragmentation. Very low-quality paths still can be selected. | +28 |
| [99] | Delay, ETX and energy | NO | Fuzzy-based | Combines the delay, ETX and energy to boost stability, reliability and energy-efficiency. | Under-specification of metrics composition (Section IV-A-2) | The enhanced stability and the slightly improved delay are not justified. Very low-quality paths still can be selected. | +20 |
| [100] [101] | ARSSI, SPRR and SRNP | NO | Fuzzy-based | Combines ARSSI, SPRR and SRNP to improve reliability with a mechanism to balance between the global quality of a path and the individual quality of its constituent links. | Under-specification of metrics composition (Section IV-A-2) | The claim that the proposed metric allows avoiding paths having low-quality links is not fully supported. It is unclear how DIOs have been incorporated into the link estimation calculation. A small number of nodes (10 nodes). | +17 |
| [105] [106] | Traffic, ETX, Data-rate, Transmit power and RE | YES | additive | Designing a new metric called ELT and using multipath forwarding for the aim of balancing the energy consumption. | Single-path routing (Section IV-A-1) | Higher risk of fragmentation. The monotonicity property is not satisfied. | +29 |
| [107] | N/A | YES | N/A | Uses multiple paths during congestion as a way of overcoming such a congestion. | Single-path routing (Section IV-A-1) | More overhead due to the new control messages. It is unclear how the congestion threshold is set. | |
| [108] | DELAY ROOT, Received packet number and ETX | YES | additive | Designing a composite multipath routing metric to mitigate congestion resulting from the sudden events in the emergency scenarios. | Single-path routing (Section IV-A-1) | Higher risk of fragmentation. No real testbed experiments. | +23 |
| [109] | ETX and RE | YES | Lexical | Design a new ETX-based and then combine it RE to improve reliability and load balancing. | Single-path routing (Section IV-A-1) | No reliability metric is used for comparison purposes. The monotonicity property is violated. The simulation tool used for evaluation is undisclosed. | +14 |
| [110] | Remaining battery voltage | NO | N/A | Introducing the remaining battery voltage as a new metric with a hysteresis of 5% to prevent excessive churn. | Under-specification of metrics composition (Section IV-A-2) | Only up to 7 nodes are used for evaluation. No justification of the higher churn experienced by OF0. | +7 |
| [111] | PD, NC, (LC) and energy | Yes | additive class-based | Combining four weighted metrics and using virtualization and SDN to supports multiple classes of traffic. | Single-path routing (Section IV-A-1) | One-hop unrealistic communication is supposed. No clarification on how DIOs are communicated in the NONSDN-based OMC-RPL. The reporting interval of the SDN-based OMC-RPL is not given. | +23 |
| [112] | BDI, PER and ETX | NO | additive | Combine BDI, PER and ETX with the focus on excluding highly depleted nodes in terms of energy. | Under-specification of metrics composition (Section IV-A-2) | The superiority of proposed OF over ETX-based OF in terms of PDR seems unjustifiable. No real testbed experiments. | +19 |

Table 4: The RPL's core operations enhancements and weaknesses.

| | The name | The module | Brief description | Limitations Addressed | Drawbacks |
|---|---|---|---|---|---|
| [115] | Memory-efficient RPL (MERPL) | Downward routes | Combining the non-storing and storing modes of operation to carry out the forwarding decisions in the *downward* direction. | Memory limitation and long source headers (Section IV-B-2, and 3) | Unclear how to set the value of the pre-specified factor N. Unpopular simulation tool is used for evaluation. |
| [116] | D-RPL | Downward routes | Using the multicast to overcome the memory limitations in the storing mode of RPL, when the node's memory overflows. | Memory limitation (Section IV-B-2) | Multicast added more complexity and sometimes it might be counter-productive. |
| [113][114] | DualMOP-RPL | Downward routes | Allows nodes operating different MOP within one physical network to understand each other and cooperate as single connected network. | Incompatible modes, memory limitation and long source headers (Section IV-B-1, 2, and 3) | Inherits the limitations of the non-storing mode in terms of higher fragmentation risk and the storing mode memory overflow. Only up to 25 nodes are used for evaluation ,not an example for a large scale-network. |
| [118] | Trickle-F | Routing maintenance | Gives the node a priority to send its scheduled DIO based on its recent history of transmission. | Suppression Mechanism Inefficiency (Section IV-C-2) | Slow convergence time due to the listen-only period. |
| [119] | Optimized-Trickle (Opt-Trickle) | Routing maintenance | Allows nodes to pick the random time, $t$, from the range *[0, Imin]* in the first interval. | Listen-only period (Section IV-C-1) | Unrealistic MAC protocol with 100% duty-cycle is used for simulation experiments. Fast convergence time, however, moderate in lossy networks as there is listen-only period in the subsequent intervals. |
| [120] | adaptive-k | Routing maintenance | Allows each node to tune its redundancy factor dynamically based on the number of its neighbors | Suppression Mechanism Inefficiency (Section IV-C-2) | The number of DIOs may not reflect correctly the number of neighbors. Slow convergence time due to the listen-only period. |
| [122] | Trickle-offset | Routing maintenance | Calculates the redundancy factor as a function of node degree. | Suppression Mechanism Inefficiency (Section IV-C-2) | Adding more a complexity by introducing two new configuration parameters. Slow convergence time due to the listen-only period. |

In [95], the authors highlighted the fact that relying on hop-count only in calculating the ranks of nodes may result in constructing paths characterized by long physical distances. As transmitter energy consumption is directly proportional to the square of the distance between communicating nodes, that may lead to routes that suffer from higher power consumption rates. The authors propose a new composite metric based on the distance between the node and its potential parent, the number of children that the potential parent has, and the hop count metric. The new framework is compared to OF0 and to the Karkazis [89] composition metric in terms of device longevity and power consumption. It is shown that the proposed framework manages to decrease significantly power consumption and enhance the longevity of the DODAG.

The instability and unreliability issues of RPL are considered in [96]. The authors report that RPL may suffer from frequent route changes that may affect network performance negatively. They assert that even though several metrics are defined for RPL, there is not a metric that represents the stability of nodes. Thus, a new stability metric, referred to as Stability Index (SI), is proposed, to overcome this issue. The new metric relies on the transmission rate of control messages to estimate the stability of links. The SI is measured at each node by adding up the weighted number of DIO, DIS and DAO control messages transmitted during a specified interval (the *Hearing Window*) and dividing the sum by the size of the interval [96]. The weighting is used to give each type of control message a different importance. The study suggests combining the new metric with ETX to boost further the protocol reliability. The proposed and combined metrics are evaluated using NS2 simulations and compared to the RPL with hop-count and ETX metrics, in terms of control message overhead, latency and packet delivery rate [96]. It is shown that the new composite metric reduces significantly the CDF of control plane overhead by up to 90% and the average number of transmissions by up to 50% compared to RPL hop count and ETX. In addition, the simulation results indicate that SI-RPL and SI-ETX-RPL outperform both ETX-RPL and HC-RPL in terms of packet delivery rate and that the amount of enhancement depends on the size of the hearing window. On the other hand, SI-RPL and SI-ETX-RPL have slightly longer latency compared to HC-RPL, as they prefer more stable and reliable paths at the cost of more hops.

### c)  Lexical-Based Composition Enhancements

A lexical composite based OF named an energy-aware objective function (EAOF) for RPL protocol is introduced in [91]. In this study, the authors highlight the issue that current RPL objective functions do not use energy-based metrics. They proposed combining the ETX metric with the residual energy of nodes in order to build a topology that is energy-efficient and reliable. A node running EAOF must first select a subset of nodes with the lowest ranks, calculated based on ETX, from its neighbors. Then, the node with the maximum residual energy is selected from this subset as a preferred parent. The parameter MAX-ETX is introduced to limit the size of the ETX-based subset from which the preferred parent is selected according the application requirements. In addition, the parameter MIN_ENER is proposed to introduce a hysteresis value when switching parents based on energy in order to ensure network stability. The study uses the popular Cooja [131] simulator with Contiki [71] to validate the proposed objective function and compare it with the ContikiRPL implementation of the ETX-based MRHOF in terms of Packet Reception Ratio (PRR), energy efficiency and network lifetime. It is shown that EAOF significantly improves the network lifetime and balances energy consumption compared to RPL MRHOF, with a negligible impact on reliability. A slight degradation in PRR is attributed to EAOF sometimes favoring balanced paths over high quality paths to enhance network lifetime.

### d)  Cross-Layer Based Composition Enhancements

A cross-layer based composition is proposed in [97], named RPL-SCSP, which combines the ETX and Queue Load aiming at providing the network with QoS support. The RPL-SCSP proposes that the selection of parent is firstly done based on the number of packets in the queue (*nqpacket*). The parent who has *nqpacket* between one and *S*, a pre-specified threshold, should be selected as the preferred parent. When multiple parents have *nqpacket* between *one* and *S,* then the selection of preferred parent is done based on the ETX values. The selection of preferred parent based on ETX values is also applied when all parents have *nqpacket* less than one or greater than *S*. It was shown by means of simulation experiments that RPL-SCSP has reduced the end-to-end delay and enhanced the network lifetime.

### e)  Average-Based Composition Enhancements

The study in [94] addresses the long single-hop problem introduced when RPL relies on a single metric such as hop count or expected transmission cost, in large networks. The authors report that, since ETX metric adds up the ETX values of the nodes along a routing path, the number of hops rather than the quality of transmission tends to have more impact on the calculated rank. Therefore, a node will tend to select the path with a small number of hops because this passes through fewer nodes and accumulates a relatively smaller total ETX [94]. Hence, the calculated ETX rank for a path with fewer hops tends to be smaller, even when such a path has constituent links with quite poor transmission quality. In a large network, a long single-hop path with a bad transmission quality can restrict the whole network affecting negatively its reliability. To overcome this problem, the study proposes combining the hop count and the ETX metrics to produce a composite metric called PER-HOP ETX. The rank is calculated based on the cumulative value of ETX along a path divided by the number of hops on that path. The new metric is evaluated using Cooja and compared with both the MRHOF and the OF0 objective functions [94]. The results indicate that PER-HOP-ETX improves PDR in dense networks while reducing power consumption and latency.

### f)  Fuzzy-Based Composition Enhancements

Several Fuzzy-based OFs for RPL have been also introduced in several studies. For instance, the authors in [98] highlighted the problem of relying on a single-metric objective function. They further pointed out that even combining two routing metrics might be insufficient to address the requirements of multiple applications as the performance objectives may vary so widely. In addition, combining just two routing metrics may enhance the network performance of the parameters associated with these, but at the expense of negatively affecting other parameters. For example, considering the ETX and latency metrics may help the RPL network to discover more reliable paths with low delay, but may lead to a battery depletion due to the overuse of some routers [98]. Thus, they assert that there is a need to design a holistic objective function that combines multiple routing metrics to optimize all significant parameters simultaneously. To fulfill this goal, they propose a fuzzy logic approach named the Fuzzy-Logic OF (FL-OF) that combines four representative routing metrics namely the hop count, node energy, link quality and end-to-end delay. It is shown that the proposed OF-FL has a tendency to reduce average hop count in comparison with the MRHOF in dense networks. In addition, OF-FL has a much better performance in terms of packet delivery ratio than OF0, and almost the same packet delivery ratio as MRHOF with ETX [98]. Furthermore, the results indicated that OF-FL has a better load distribution among nodes leading to a more balanced energy consumption than F0 or MRHOF with ETX. Finally, OF-FL demonstrates the lowest average end-to-end delay for nodes on the edge of network while the delay is comparable with the standardized OFs in other cases. However, OF-FL experiences higher churn compared to MRHOF with ETX [98].

Another fuzzy-based approach to combining routing metrics is introduced in [99]. The authors used a two-stage fuzzy process to combine three linguistic variables (routing metrics), namely, delay, ETX and energy. In the first stage, the delay and ETX are combined to compute what they call Quality of Service (QoS). In the second stage, the energy is combined with the computed QoS value. The proposed fuzzy-based approach is then evaluated against ETX-RPL using a real testbed network of twenty-eight sensor nodes. The two protocols are compared in terms of packet loss ratio, energy consumption and routing stability (number of preferred parent changes). It is reported [99] that the fuzzy-based approach outperforms ETX-RPL in terms of packet loss ratio by up to 20% and slightly enhances end-to-end delay. In addition, the proposed approach is shown to build a topology of more stable routes with an average of 6.63 parents change per hour compared to ETX-RPL with an average of 43.52.

A third fuzzy-based routing metric is proposed in [100] [101] referred to as Opt-FLQE$_{RM}$. This composite metric considers three link estimation metrics: Average Received Signal Strength Indicator (ARSSI), Smoothed Packet Reception Ratio (SPRR), and Smoothed Required Number of Packet retransmissions (SRNP). These three routing metrics are combined using a fuzzy approach that produces a score from the range [0...100], where 100 is the best quality and 0 is the worst. To select the optimal path, the inverses of the individual link qualities are added and then the path with minimum value is selected. The authors claim that relying on the inverse when selecting the optimal path allows the metric to avoid low-quality links while favoring paths with fewer hops. For evaluation purposes, the proposed routing metric is compared to RPL, ETX-RPL and the four-bit CTP [102] metrics using the well-known Cooja simulator in terms of average packet loss, average end-to-end delay, average hop-count and average power consumption. The authors show that their proposal produces the lowest packet loss, the lowest end-to-end delay and the lowest churn (number of parent changes) among the compared metrics. The superiority of Opt-FLQE$_{RM}$ over ETX is attributed to the conservative approach used by ETX to estimate link quality: this is based on data traffic, which is only obtained after topology establishment. In contrast, Opt-FLQE$_{RM}$ bases its calculation for link qualities on both control and data traffic resulting in an accurate estimation of link quality at the time of topology construction that results in constructing more stable paths [100].

**Summary and Insights:** The study in [89] was the first attempt to provide RPL with a way to quantify routing metrics and allow them to be combined lexically or in an additive manner. Although the study presents a good proposal to distribute energy load among nodes, by combining the RE and the hop-count, it does not elaborate on the effect of this combination on the network reliability, a critical performance criterion. It is also unclear whether the study uses the aggregated value of the RE metric or a local optimum value. A major issue with the study in [90] is that only up to six nodes are used for the simulation experiments, which may be insufficient to reach the conclusions reported. In addition, the authors did not elaborate on how the composite metric may affect the reliability of the network.

The shortcomings of the articles in [89][90] are addressed in [91]. First, the author introduces the parameter MIN_ENER to limit the churn in the network due to energy-related parent switches. Second, the study introduces a reliability-related performance evaluation of the composite matric. However, only 25 nodes were used in the simulation experiments, which means that conclusions reached cannot be generalized to larger networks.

Although the study in [92] claims that the gain in network lifetime is obtained without affecting network reliability, the study does not reports any results regarding the reliability nor does it justify how the authors reach this conclusion. In addition, the authors used their own bespoke simulator for evaluation purposes, which may lack in features compared to the well-known simulators such as Cooja. The study reports setting the Trickle timer interval for emitting DIOs to 1 hour. It seems the authors have configured only one interval in their simulations, which is a confusing deviation from the normal operation of Trickle protocol.

In [93] there is a higher risk of layer 2 fragmentation as DIOs transmitted by nodes running SCAOF need to carry a relatively large poll of parameters in their headers. This represents a serious problem in the LLNs as it increases the probability of errors and packet loss, especially in multipath routing.

A major issue with PER-HOP ETX metric proposed in [94] is that the monotonicity property of the combined metric is not satisfied so the network might be at the risk of forming loops. The work in [95] suffers from the problem that the estimation of a node's positions in real testbed deployments is not a straightforward process and so live physical distance estimations are likely to be either imprecise (e.g. RSSI) or power-hungry (e.g. GPS) [124][125].

The frequency of control messages (DIOs, DAOs, and DISs) is used in [96] to measure stability of the node and the routing topology but, in some cases, the higher frequency of control messages does not imply higher instability. For instance, a node with a higher number of children will have to transmit a higher number of DAOs than a node with a fewer number. In this scenario, it is clear that the number of children has caused the higher control overhead, and not the instability problem. A more elegant solution is to base the measuring of the instability index on the DIO messages alone.

Finally, the fuzzy-based approaches are known to incur greater complexity compared to other approaches, especially when multiple instances exist under the same RPL topology [77]. For instance, in [98] more than four parameters need to be transmitted within the DIO metric container. Thus, there is a higher risk of fragmentation, which incurs more overhead due to the larger size of DIOs [77]. In [99] the stability of routes is claimed to be the reason of the superiority of proposed approach; however, no justification is given to explain why the fuzzy-based approach is more stable. The lack of justification also applies to the slightly improved delay. The work in [100] does not clarify how the control traffic messages (DIOs) have been incorporated into the link estimation calculation. Finally, Opt-FLQE$_{RM}$ tends to favor shorter paths in terms of hop count, which may result in selecting paths containing low-quality single-hop links.

### 2) OF Enhancements Based on Multi-path Routing

In order to overcome some performance issues resulting from single-path based routing in RPL, several multipath forwarding optimizations have been proposed and still other studies have proposed multi-path forwarding approaches that use composite metrics. For instance, the authors in [103] propose a probability-based load-balancing multi-path solution for RPL referred to as LB-RPL. LB-RPL achieves load balancing by having each node distributes traffic among its top $k$ parents, in terms of *rank*, based on their traffic load. A parent experiencing heavy load may signal its status by delaying the broadcasting of its scheduled DIO message. This enables child nodes to remove that parent from their top $k$ and hence, exclude it from further data forwarding. It is shown [103] by means of simulations that LB-RPL outperforms RPL in terms of packet delivery ratio, delay, and workload distribution.

The work in [105][106] highlights the advantages of incorporating multipath forwarding schemes into the RPL protocol. Intuitively, the multipath mechanisms have been proven to have a wide spectrum of benefits such as improving fault-tolerance, enhancing reliability, minimizing congestion and improving QoS. The authors propose a multi-path routing mechanism based on RPL in order to allow the protocol to forward traffic to multiple preferred parents. The study asserts that a routing metric must: (1) capture the variations in link quality; (2) use energy-efficient paths to maximize the end-to-end reliability; and (3) minimize the energy expenditure for those nodes consuming the most energy (the bottleneck nodes).

In this regard, a new metric is proposed, referred to as the Expected Lifetime metric (ELT) that aims to balance energy consumption among network nodes and maximize the lifetime of the bottleneck nodes. The network lifetime is defined as the time before the first node dies (runs out of energy). The ELT of a specific node is calculated by: (1) computing the throughput of that node based on its own traffic and also the traffic of its children; (2) multiplying the average number of retransmission by the calculated traffic; (3) computing the time ratio required for transmission based on the sending data rate; (4) computing the energy consumption based on the transmission power of the radio only; and, finally, (5) calculating the ELT as the ratio between the node's remaining energy and the energy calculated in the previous step. Based on the ELT calculated value, the bottleneck nodes are first identified and advertised along the topology, then a multiple-parents, energy-balanced topology is constructed, in which the traffic is balanced among parents with careful consideration of bottleneck nodes [105][106]. The proposed protocol is evaluated using WSNet [129] and compared to RPL considering the metrics of: residual energy, the ETX-using-hysteresis objective function, and a linearly combined metric of ETX and residual energy. The experimental results indicate that the proposed multipath ELT has almost the same reliability as ETX although selecting the paths with maximum residual energy. Multipath ELT was also found to enhance routing stability through preventing sudden changes in the parent weight in comparison with standard RPL [105].

In [107], the authors again highlighted the issue of RPL being single path routing protocol and the incapacity of standard objective functions to provide multipath routing. The ultimate goal of the study is to provide RPL with multipath routing capabilities that will enable the protocol to react efficiently to the congestion [107]. Thus, the authors propose an extension referred to as a multi-path RPL (M-RPL) that provides temporary multiple paths during congestion. In M-RPL, the packet delivery ratio (PDR) is used by the forwarding nodes to detect congestion. If a forwarding node on a routing path detects that the PDR has decreased below a specific threshold, the node sends a notification to its children, by means of DIO messages, informing them of congestion. Each child node that hears the congestion advertisement message starts multipath routing by splitting its forwarding rate in half. Thereafter, only every second packet is sent to its original congested parent while the others are forwarded to any other parent from its parent list [107]. The proposed protocol is evaluated using Cooja and compared to RPL with MRHOF in terms of energy consumption, latency, and throughput. Their simulation results show that M-RPL has better throughput and lower per-bit energy consumption than RPL, due its splitting mechanism. The results also indicated that while the delay of M-RPL is initially comparable to RPL, this changes when congestion begins. Initially M-RPL experiences greater delay as multiple paths are introduced but when the network stabilizes; M-RPL gets to outperform RPL in terms of delay [107].

The work in [108] proposes a multi-path forwarding approach based on a composite metric. The authors point out that the two single-metric RPL's objective functions are vulnerable in scenarios where a sudden increase in traffic volume introduces a congestion, resulting in significant delay and packet loss. The authors propose a congestion avoidance multipath routing protocol, referred to as CA-RPL, whose primary goal is to enable the network to react quickly and reliably to sudden events. They have designed a composite routing metric based on the ContikiMac duty cycle protocol with the aim of minimizing the average delay towards the DODAG root, referred to as DELAY ROOT. Under this metric, a node saves time by first learning the wakeup phase of its candidate parents and then sending the packets to the first awake parent [108]. CA-RPL is a composite multi-path routing metric that combines the new proposed DELAY ROOT with the number of received packets and ETX to calculate the path weights. Cooja with Contiki operating system is used to compare the proposed protocol with standard RPL in terms of latency, packet loss ratio, throughput and the packet reception number (PRN) of the DODAG root per unit time. The experimental results illustrate that the proposed protocol relieves network congestion and enhancing the PRN by up to 50%, the throughput by up to 34%, packet loss by up to 25%, and average delay by 30% compared to RPL.

The authors in [109] reported that the ETX metric used in RPL is inefficient in quantifying the quality of links as it only "*reflects the quality of a single link*". To overcome this issue, the study proposes a link quality aware routing protocol for LLNs referred to as LQA-RPL. LQA-RPL calculates the rank of a node based on the quality of links to all its neighbors, which is derived from the ETX and defined as the expected probability of unsuccessful transmissions. If a node has more than one parent in its parent set, the node uses multi-path routing by selecting the parent with the maximum residual energy to act as next-hop relay node to the DODAG root [109]. LQA-RPL is evaluated and compared to RPL with hop count in terms of packet delivery ratio, energy consumption, and network lifetime. The reported results indicated that LQA-RPL outperforms RPL in terms of PDR, which is attributed to the higher number of candidate parents. It is also shown [109] that LQA-RPL can balance energy consumption due to its capacity to distribute the traffic among multiple candidate parents based on residual energy, prolonging the network lifetime.

The work in [110] has reported a new energy-based OF that proposes selecting the preferred parent based on the remaining energy with a hysteresis value of 5% to reduce network frequent changes. The remaining energy is obtained by polling each node to check its battery voltage, which is claimed to be a good indicator of remaining energy [110]. The proposed OF is evaluated using two testbed deployments (basic and extended), and compared to RPL objective functions (OF0, MRHOF) in terms of packet loss, delay, energy consumption and network churn. In both testbeds, a sink and seven sensor nodes are deployed in a building area and differentiated by node position, distance between nodes, RF interference and noise. Based on the obtained results, it is shown that the proposed OF has improved network lifetime by up to 40% compared to RPL's OFs. The authors also claim that their OF lowered the delay in the Basic Deployment compared to MRHOF, which was expected, and to OF0, which was not. In explaining the latter case, they observe that OF0 suffers from excessive churn and frequent changes in the network topology resulting in higher delays and more energy expenditure. Pertaining to packet loss, the reported results show that the proposed energy-based OF is superior to RPL OF0 but is outperformed by RPL MRHOF.

The authors in [111] propose an optimization for RPL referred to as Optimized Multi-Class RPL (OMC-RPL) based on virtualization and software-defined networking techniques. The study asserts that standard RPL faces two significant issues when offering QoS. The first is the absence of a holistic and comprehensive objective function. For example, an objective function may enhance delay but at the cost of higher energy consumption as all packets serve the same paths with the minimum delay. The second issue is that RPL does not support a mechanism for data classification, which is critical component in ensuring the QoS [111]. Thus, a holistic objective function that supports multiple data classes is needed [111]. The steps of OMC-RPL are as follows: first, the nodes send the information required to construct the virtual DODAG to the SDN controller, using one-hop communication; then the SDN controller calculates the ranks of nodes in the network for each traffic class using a custom weighted-metric objective function [111]. The main parameters of the proposed objective function are the Propagation Delay (PD), Node Congestion (NC) and Link Congestion (LC). Energy is considered as a secondary parameter and is thus incorporated into the objective function in a way that it can be removed or considered as desired [111]. The weight values of the objective function parameters were found using the Particle Swarm Optimization (PSO) algorithm. OMC-RPL is simulated with four different classes of traffic and compared to standard ETX-RPL in terms of end-to-end delay, packet loss, network lifetime and traffic overhead. OMC-RPL then outperforms RPL in terms of end-to-end delay for the class of traffic that requires minimum delay and likewise performs better than RPL in terms of PDR with the class of traffic that requires reliability. It is also found [111] that OMC-RPL reacts better to network failures since it can use a backup parent to replace a failed one. OMC-RPL outperforms RPL in terms of network lifetime by up to 41% and shows a better fairness in energy distribution by about 18%. The study also reports that incorporating the SDN controller with OMC-RPL reduces the number of exchanged control packets compared to both OMC-RPL and standard RPL by approximately 62% and minimizes the energy consumption by more than 50% compared to standard RPL [111].

In [112], the authors propose a new composite energy-aware routing metric, $RER_{BDI}$, which aims at enhancing the energy consumption of LLN nodes. The study considers the Battery Discharge Index (BDI) and the Residual Energy Ratio (RER) of nodes for taking the routing decision. The study also defines a new objective function, referred to as $OF_{RBE}$, which combines the new proposed metric with ETX for calculating the rank and selecting the preferred parent. The study mentioned that using RER as a primary routing metric favors paths with a higher average residual energy. BDI was introduced as an additional cost function to favor paths that do not include nodes whose battery energies have been depleted or overburdened nodes [112]. Hence, the protocol avoids selecting paths in which some nodes have low residual energy even though the average residual energy is high. The Cooja simulator is used to compare the proposed scheme to standard RPL in terms of PDR, network lifetime and energy consumption. It is found that the RER metric outperforms ETX in terms of energy consumption by avoiding paths with lower average residual energy. However, there is still a chance that nodes with a low power profile could be selected as preferred parents, which may decrease network lifetime [112]. This situation is improved in $RER_{BDI}$, which favors paths with higher average residual energy, while avoiding ones that include nodes with very low energy [112]. The new metric enhances network lifetime compared to RPL with hop-count, ETX and RER, but is slightly outperformed in terms of PDR, as it does not consider link quality [112]. Finally, it was shown that the proposed objective function has exceled at both the network lifetime and the PDR compared to RPL with ETX and Hop count.

**Summary and Insights:** The implicit signaling through delayed DIO proposed in [103] has no extra overhead, however, a lost DIO might easily be misinterpreted as delayed, giving a false indication of higher workload at some nodes. In addition, the long transmission periods of Trickle's DIOs may cause slow recovery. Moreover, the protocol may suffer from the herding effect problem by always changing parent set members [104] .In [106], because several parameters must be exchanged (i.e. data rate, retransmission count, throughput, transmission power and residual energy) to calculate the rank, this approach increases the size of DIO messages increasing the risk of fragmentation. This represents a problem in LLNs when multipath routing is used as two fragments belonging to the same packet may take different paths, increasing the probability of errors and packet loss. In addition, the monotonicity property does not hold for the ELT metric; hence, the study proposes to use ETX to build the DODAG and the ELT to calculate the rank of nodes. This would introduce an extra complexity to an already complex protocol [77].

The study in [107] suggests that each child node must report its current forwarding rate to its parent node by means of DAO messages to calculate the PDR. Apart from being an optional feature in RPL, only used when downward paths are needed, DAO messages are costly in term of overhead and energy consumption as they are transmitted in end-to-end fashion. The proposed protocol in [108] is based on ContikiMac that assumes that all nodes have similar wakeup intervals, which may not hold in all LLNs scenarios [77]. In addition, many additional fields are carried in the DIO message, which increases the risk of fragmentation. An obvious issue with the proposed protocol in [109] is that it is compared to RPL with hop count, even though the problem statement focuses on explaining the unsuitability of ETX metric to quantify the reliability of links. Thus, it would seem more logical to compare the proposed combined metric with ETX as both quantify link reliability [77]. Another issue is that the reported metric, if implemented according to the algorithm shown in the study, would violate the monotonicity property of rank, potentially resulting in a loop-prone DODAG topology, an issue that must be avoided.

A noticeable issue related to the study in [110] is that only seven nodes are used for the evaluation which is very small to display the advantages of the proposed OF. In addition, while the increased delay in OF0 was attributed to the high churn, no justification was provided as to explain why OF0 experiences a higher churn given the perceived stability of the hop-count metric.

In [111], it is assumed that all nodes are within the range of the SDN controller so that the messages can be communicated by one-hop; however, this is unrealistic in the majority of cases. In addition, while RPL uses a Trickle timer for communicating control packets, it is unclear what mechanism is used by the NONSDN-based OMC-RPL (OMC-RPL without SDN controller) for communicating such messages. Furthermore, even for SDN-based OMC-RPL, the reporting interval to the SDN is not quoted, although it could have a big effect on the control plane overhead.

The decadic logarithm (i.e. log with base 10) is proposed in [112] to calculate the BDI (an additional cost inversely proportional to the node's residual energy). The calculated additional cost, based on the node's initial and residual energies, will be very small compared to the node's residual energy so will have no significant effect on the composite metric's final cost. Although, the study suggests using different weights to adjust the influence of metrics involved, restricting the weights to be within the interval [0, 1] limits the extent to which the influence of BDI can be adjusted.

### 3) Common Challenges and Pitfalls of OFs Enhancements

Although combining two or more metrics may give an application the capacity to optimize more than one aspect at a time, it may lead to undesirable consequences if not designed efficiently. Several problems have been identified as follow.

Firstly, using multiple metrics means that a higher load of information need to be carried in DIO control messages, which in turns increases the risk of layer 2 fragmentation [77]. Apart from consuming network resources such as energy and bandwidth, fragmentation represents a serious problem in the LLNs especially when multipath routing is used, as two fragments belonging to the same packet may take different paths, increasing the probability of errors and packet loss [77]. The risk of fragmentation is more evident in fuzzy-based approaches as they feature greater complexity, especially when multiple instances exist in the same RPL topology. Secondly, in weighted composition, it is usually hard to decide on what weights to assign to the component metrics and whether the assigned values should be static or dynamic according to the context (e.g. time, position). Thirdly, some suggested metrics, such as node position, cannot be easily estimated in real environments [124][125]. Fourthly, the composite metric may fall into the trap of giving one metric such a high priority that it behaves effectively as if it was itself a single metric. Finally, designing a composite metric that violates the monotonicity property should be avoided as it can lead to loops that harm reliability and waste resources [89]. In fact, the monotonicity property should be preserved even with single-based metric OFs.

Unrealistic assumptions related to the operations of the protocol itself or its perceived environment is another pitfall that may lead to false conclusions. Such assumptions include one-hop communication range, building network-wide decisions based on optional features in the standard, non-duty-cycle and synchronous MAC protocols. For instance, in [96], the frequency of control messages was used to measure the stability of the node and the routing topology but the higher frequency of control messages does not necessarily imply higher instability and this may mislead the routing decision.

The RPL standard is intended to run on LLNs encompassing thousands of sensor nodes; however, a number of the surveyed enhancements were evaluated on networks comprising less than 10 nodes (e.g., only six nodes were used in [90]). The small scale of the test network is inadequate to reach strong conclusions reported or display the advantages of proposed enhancements, as results cannot be generalized to large-scale deployments.

Multi-path routing techniques are highly desirable in LLN environments as they have been proven to provide a wide spectrum of benefits such as improving fault-tolerance, enhancing reliability, minimizing congestion, increasing network capacity (bandwidth aggregation) and improving QoS [126]. However, multi-path routing techniques do have their own disadvantages that should be considered carefully when designing routing primitives for LLNs [126][127]. One of the primary concerns associated is that multi-path approaches introduce greater complexity and overhead. In such techniques, the intermediate nodes are required to maintain the state of multiple routes to a destination; this might be infeasible for memory-constrained LLN devices especially in the case of downward traffic, where a node must store routing entries for all destinations in its sub-DODAG [127]. The way the data packets are allocated to multiple paths represents another challenge [127]. When fragmentation occurs, fragments of the same packet might be transmitted on different routes raising the need for packet re-ordering. This risk is high if a round-robin traffic allocation is used to distribute traffic among multiple paths based on per-packet granularity [127].

Ensuring full efficiency of multipath routing requires the discovery and the maintenance of network-wide node-disjoint paths, which creates an extra overhead and may be infeasible in resource-constrained networks with highly dynamic links and scarce energy resources [128]. Moreover, the broadcast nature of the wireless medium may impede the goals of reducing congestion or load balancing due to the *route-coupling effect*, a phenomenon in the wireless medium that takes place when several paths are located in close proximity causing communication interference and increasing the risk of collisions [128]. Although location-aware routing can be used to mitigate the effect of route-coupling problem by constructing non-interfering routes, the high overhead incurred by such techniques in terms of computational and communication complexity makes them unsuitable for the resource-constrained LLN devices [126].

### B. Routing Maintenance Enhancements

Several extensions have been proposed to overcome the problems associated with introducing the listen-only period and the suppression mechanism inefficiency in RPL's routing maintenance primitive as detailed below.

#### 1) Suppression Oriented Enhancements

The first trial to solve the Trickle issues in LLN routing is the study in [118]. The study reports that suppressing RPL control messages by means of the Trickle algorithm may result in constructing sub-optimal paths, worsening as the number of suppressed DIOs increases. This behavior is explained by the fact that Trickle is originally designed for propagating the same piece of information with the least number of messages across a network [118]. However, the DIO messages in RPL are not necessarily identical as the information carried strictly depends on the source of the message; thus, suppressing one or another is not always equivalent [118].

To address this issue, an enhanced version of Trickle referred to as Trickle-F [118], which strives to guarantee a fair multicast suppression among RPL nodes is proposed. Trickle-F gives each node a priority to send a scheduled DIO based on its suppression history. The more the node suppresses DIOs, the higher the chance it will transmit in the next interval frame. The proposed enhancement is compared to the original Trickle under RPL by means of simulation in terms of network stretch, average energy consumption and the distribution of suppressed messages. It is shown [118] that Trickle-F reduces the number of nodes with sub-optimal routes compared to Trickle while displaying the same energy consumption profile. This superiority is attributed to the spatial fairness achieved by Trickle-F among nodes.

The work in [120] highlighted the ambiguity associated with configuring the redundancy parameter, $k$, in RPL networks. For instance, the Trickle RFC [69] states that typical values for $k$ are 1-5, whereas the RPL RFC [14] sets the value 10 as the default value. However, the best value for the redundancy constant is claimed to be between 3 and 5 in the last IETF draft titled "*Recommendations for Efficient Implementation of RPL*" [121]. Finally, it is recommended in the MPL RFC [117] to set the default value of $k$ to one. This shows that the optimal setting of $k$ is not a trivial task and relies greatly on the application scenario as well as the network topology at hand [120]. The authors here suggest setting $k$ for each node individually based on that node's degree, a mechanism they call *adaptive-k*. They use the number of Trickle messages received during a specific interval as an implicit indication of node degree. By means of simulations and testbed experiments, it is shown that the proposal improves the performance of RPL through lowering the control-plane overhead while enabling the discovery of more optimal routes.

In [122], it has been shown by means of a mathematical analysis that the single redundancy constant adopted by Trickle may result in higher transmission load and consequently higher power consumption rates for those nodes having fewer neighbors. To alleviate this issue, the study proposes an enhancement of Trickle in which each node calculates its own version of the redundancy constant as function of its degree. Each node having a number of neighbors less than a pre-specified threshold, called the *offset*, will set its redundancy constant to *one*. The redundancy constant of other nodes should be set by subtracting the number of neighbors from the offset and taking the ceiling of dividing the result by another predetermined value called the *step*. It is shown by simulations that the proposed algorithm balances the transmission distribution among network nodes in comparison with standard Trickle [122].

#### 2) Listen Interval Oriented Enhancements

In [119], the authors highlighted the problem of increased latency resulting from introducing the listen-only period in the Trickle algorithm. To address this problem, an Optimized-Trickle, (Opt-Trickle) is proposed. The authors observe that nodes receiving inconsistent transmissions simultaneously will reset their timers (returning to $I_{min}$) immediately, thus exhibiting a form of implicit synchronization. Such a synchronization in fact eliminates the need for the fixed listen-only period in the first interval and allows the affected nodes to pick a random time, $t$, from the range [0, $I_{min}$]. This is the only modification in Opt-Trickle.

**Summary and Insights:** Although Trickle-F [118] has succeeded to some extent in solving the sub-optimality of constructed routes; the algorithm still suffers from slow convergence time due to the listen-only period. The study pertaining to Opt-Trickle [119] assumes a MAC protocol with 100% duty-cycle, which is neither reasonable nor realistic. Furthermore, Opt-Trickle still has a listen-only period in subsequent intervals, which would contribute to the increased latency especially in a lossy network where it is not guaranteed that a transmitted multicast message will reach all of its destinations at its first transmission in the first interval. In [120], it is unclear why the study resorts to the number of messages received at specific node and not the number of actual neighbors, to estimate indirectly the network density at that node. Although this method might give an approximately accurate estimation for node degree when the network is characterized by synchronized intervals among its nodes, it may suffer from inaccurate estimation in non-synchronized networks. For instance, in a non-synchronized network, the frequency of transmission may differ significantly from a node currently in its minimal interval to another node currently in its maximum interval. The node in its minimum interval will transmit more frequently, giving the receiver node an impression that it has more neighbors than it does actually has, thus affecting the accuracy of the network density estimation. The work in [122] did not demonstrate the impact of the proposed enhancement either on the quality of constructed routes nor on network power consumption. In addition, introducing two new parameters, the step and the offset, adds complexity, which is best avoided.

### 3) Routing Discovery and Maintenance Key Challenges

In general, there are two routing discovery maintenance schemes in the context of LLNs; proactive and reactive [132][133]. In proactive routing, the process of establishing routes is carried out in advance and maintained periodically, which induces a large amount of overhead, although with minimal forwarding delay [132]. On the other hand, in reactive routing, the routes are only discovered when needed, thus suffering from higher delay compared to proactive schemes [132]. Reactive routing schemes have been the preferred option when the network features a high number of mobile nodes whereas proactive schemes are preferred in stationary networks (e.g. LLNs). However, the resource-constrained nature of these networks imposes several challenges on proactive route maintenance. Despite the stationary nature of the majority of scenarios, LLNs do exhibit some dynamicity that may render the network unstable, dictating the need for a rapid and reactive response [54][55][56][57]. Choosing small update intervals has the advantage of faster propagation, but with a high communication overhead; long update intervals, on the other hand, have lower communication overhead, but disseminate routing information slowly [53]. To address these problems, Trickle [53] has introduced the notion of adaptive and dynamic interval size. The idea is to start propagating routing information at a high transmission rate then gradually reduce when the network reaches its steady phase, ensuring rapid propagation and low overhead. Trickle uses the term *inconsistency* to describe the point of time at which the network must start transmitting at its fastest rate. Although Trickle and its extensions have defined how information exchange should proceed in the consistent and inconsistent sates of the network, they fail to define clearly what constitutes inconsistent or consistent transmission in the context of routing. We argue that an efficient routing maintenance should clearly define what constitutes consistent and inconsistent state and define corresponding detection mechanisms.

### C. RPL Downward Routes Enhancements

Some effort has been directed at increasing the efficiency of constructing *downward* routes based on combining both modes of operation (hybrid mode), or using multicast techniques.

### 1) Hybrid Based Enhancements

For instance, the issue of interoperability between RPL's non-storing and storing modes of operation has been highlighted in [113] [114]. To solve the interoperability problem, the authors propose DualMOP-RPL [114], which allows nodes operating in different modes to understand each other and cooperate as a single connected network. In this regard, two major enhancements are suggested on the top of RPL: firstly, nodes operating in storing mode should attach source routing headers to transmitted messages so that nodes configured in non-storing mode can understand them; secondly, all nodes operating in non-storing mode should advertise their destination prefixes in a hop-by-hop manner rather than the end-to-end approach currently specified [113][114]. Hence, a router configured in storing mode is able to store the routing information of all other nodes in its sub-DODAG, even when some of its children are configured in non-storing mode. DualMOP-RPL [114] is evaluated using both simulations and testbed experiments and compared to RPL in terms of end-to-end packet reception ratio (PRR) ratio and is shown to outperform RPL in terms of PRR when the two modes of operation are mixed together in a single network consisting of 25 nodes. This due to mixed-mode network partitioning in RPL and resultant selection of non-optimal paths.

The authors in [115] aim to mitigating the issue of storage limitation in storing mode. They note that RPL storing mode requires every node to maintain the routing state of all other nodes in its sub-DODAG, and many nodes, especially those close to the root, may not have adequate resources for this. To overcome this issue, the authors propose *memory-efficient* RPL (MERPL) [115]. The primary idea here is that a node, whose routing entries reach a pre-specified threshold N, should delegate a child in its sub-DODAG to act as its store. The overloaded node should then remove from its routing table all routing entries whose next hop is that delegated child. Next, all those destinations reachable through the delegated child should be advertised to the DODAG root in a separate DAO. A hybrid approach of non-storing and storing modes of operation is employed by the network nodes to carry out the forwarding decisions in the *downward* direction. To validate MERPL, it is compared to standard RPL in terms of the average number of routing table entries, average path length, and the number of items in the source root. A Python language simulator is used with network sizes of 576 and 1204 nodes [115]. The results show that MERPL does indeed reduce the routing entry storage requirements especially at nodes near the root. The average number of items in a source route is reduced by 61.5% compared to RPL when N is set to 10. MERPL average path length is also shown to be shorter than that in RPL in non-storing mode, but slightly longer than in storing mode.

### 2) Multicast Based Enhancements

A different approach for overcoming storage limitation in RPL storing mode is reported in [116]. Here it is noted that when a node fails to store a new destination routing entry, it should not propagate the information further, as it will not be able to forward to that destination. A negative effect of this behavior is that a path is partially built but is useless since the destination is unreachable by routers higher in the DODAG, including the root [116]. To address this problem, the authors suggest *D-RPL* [116], which integrates multicast dissemination into RPL storing mode. Here any node that fails to announce a destination, either for one of its children or for itself, should first register itself with a special multicast group. Then, the multicast address of this special group can be used by the DODAG root to communicate with such destinations unreachable through normal operation. The multicast can be implemented by any suitable protocol such as MPL (Multicast Protocol for Low power and Lossy Networks) [117] or via the multicast mechanism in the RPL protocol itself [116]. D-RPL is evaluated by Cooja with Contiki and compared to the standard RPL in terms of PDR, radio duty cycle and the end-to-end delay. The simulation results show that D-RPL yields significantly better performance in terms of PDR with a 6-fold improvement compared to ContikiRPL. Both protocols have comparable performance in terms of average duty cycle when the number of nodes is less than 60 but above this size, D-RPL has a higher average duty-cycle due to its higher delivery rates. The average end-to-end delay also increases in D-RPL compared to RPL, but this is attributed to the forwarding mechanism in SMFR that opts to delay packet forwarding at each hop for a specific time to avoid collisions. Finally, it is concluded that there is a higher cost in terms of delay and average duty cycle to deliver packets using D-RPL, "*but this cost is only paid for packets that would otherwise not be delivered at all*" [116].

**Summary and Insights**: The authors in [114] use only 25 nodes for evaluation purposes, which is insufficient to prove the superior performance of DualMOP-RPL in large scale-networks. In addition, although enabling interoperability between RPL modes enhances performance, problems still occur due to the limitations of the two modes themselves. For instance, enabling interoperability does not solve the issue of long source headers in the non-storing mode, nor that of memory-overflow in storing mode. In [115], other than the number of nodes, no other simulation parameters are reported and in particular there is no clear specification on to how the value of *N* should be set.

In [116], although it is claimed that the additional cost "*is only*

paid for packets that would otherwise not be delivered at all*", this may only hold true if we assume that all node routing tables overflow at the same time. It is not the case if the nodes experience overflow at different times because such nodes then flood the network with multicast packets, negatively affecting the flow of data from all nodes, including those not currently experiencing overflow. The negative affect is thus not limited to packets that would otherwise not be delivered at all.

### 3) Downward Routes Enhancements Common Challenges

The development of efficient *downward* routing schemes for RPL faces two primary challenges: firstly, memory limitations constrain the number of entries that can be stored in a node's routing table; secondly, packet size is limited by the underlying communication technology. The first restricts the number of nodes that can be accessed by the root (table-driven) while the second limits the number of hops that can be inserted into the IP packet header (source routing) [113] [114][116]. This is exacerbated by the fact that RPL is a single-path routing protocol which prevents nodes from benefiting from the combined capabilities of multiple parents [105][107]. Although [114] propose combining the table-driven approach (storing mode) and the source routing approach (non-storing mode), both are subject to the storage and hop constraints. As discussed above, the use of multicasting proposed in [116] may only be beneficial in limited cases. In other cases, using multicast will be just counterproductive, as it will harm the normal traffic efficiency of those un-overflowed nodes.

## VI. RESEARCH DIRECTIONS AND OPEN ISSUES

Having discussed RPL limitations and research efforts to address these, we are now in a position that enables us to answer the research question of this survey "*have the research attempts to address the limitations of RPL succeeded, or does more need to be done?*" In this regard, we articulate that attempts to address RPL limitations have mostly fallen short in tackling such limitations and there are still a need for further research efforts. The following are some directions not fully addressed in the literature that might take RPL further.

### A. Downward Traffic Patterns

As discussed previously, RPL did not pay much attention to the optimization of *downward* traffic (P2MP) compared to that of *upward* traffic (MP2P) giving rise to several limitations. While the issue of interoperability between RPL's two modes of operation was addressed successfully in [113][114], the issues of memory limitations in storing mode of operation and of long source headers in non-storing mode have not been fully addressed limiting the scalability of the protocol and hindering its usability in large-scale bi-directional deployments. We argue that there is still a need to develop solutions to overcome these limitations. Further, the under-specification of DAO and DAO-ACK control messages has not been tackled also, although this has a significant impact on the efficiency of the protocol and the timing of these remains an open question [116].

## B. Load-Balancing

While there has been much research into developing load-balancing primitives, it is also evident, as reported in [104], that the majority of the proposed mechanisms suffer from instability caused by the frequent switching of preferred parents, an issue that may undermine the benefits of load-balancing. Hence, we articulate that an efficient load-balancing mechanism must not only distribute traffic among nodes, but must also ensure stability. Thus, a solution that considers such an issue (i.e. providing load balancing while maintaining the stability) is yet to be developed, as all evaluated extensions are not able to address this concern efficiently.

## C. Metric Composition

We understand the need for combining more than one metric to achieve more than one application goal. However, a number of factors are to be taken into consideration when designing a composite metric. First, the number of involved metrics should be kept minimal, as a large number of combined metrics may only be counterproductive. This is because that the more number of combined metrics, the larger is the size of DIOs, which increases the risk of fragmentations and routing errors [77]. Second, for additive composition, the process of assigning weights for combined metrics becomes more difficult as the number of metrics increases. Hence, we advise that lexical composition should be preferred over additive composition. Third, cross-layer routing design, where multiple metrics from different layers are combined, seems to provide more efficient solutions for routing in LLNs [97] [104]. The cross-layer approach can be utilized to satisfy different requirements such as load-balancing [104] and energy-efficiency [97].

## D. Single-instance vs Multi-instance Optimization

One of the design considerations of RPL is that multiple applications with conflict routing requirements may run concurrently in a single physical topology. This can lead to multiple instances, where each has its own OF (i.e. one OF per instance) featuring one or more routing metrics. However, the simultaneous operation of multiple instances will increase the implementation and configuration complexity of the protocol and RPL's specification does not provide a guidance on how this should be done [18] [77].

Some recent studies such as [18][77] propose removing this feature from the protocol but, while this may overcome the problems of interoperability and implementation complexity, it may hinder the capacity of the protocol to accommodate antagonistic requirements within a single network. We argue here that it is too early to judge whether the multi-instance feature of RPL should be removed or not, especially in the absence of research studies that evaluate and compare both scenarios. However, before such evaluation can be carried out, it is necessary for the standard to specify the operation of multiple instances so the researchers can avoid unrealistic assumptions that may harm the credibility of any conclusion reached.

## E. Real Testbeds Evaluation

In this survey, we have reviewed and analyzed a significant number of proposals for RPL's enhancements that raises some observations regarding the use of testbeds evaluation in such proposals. An interesting observation is that only about 10% of reviewed articles have used real testbeds for the evaluation purposes. Even more interesting, the maximum number of nodes in such testbeds has never been greater than 30, a size only applicable to home automation applications. Simulation-based evaluation, while useful, cannot fully reflect all aspects of real scenarios. We argue therefore that research efforts to overcome the limitations of RPL should aim to evaluate solutions using real large-scale testbeds to reflect the settings in real-world environments. Such a large-scale open testbed for IoT research is available in [130] that provides an infrastructure populated with numerous IoT wireless sensor devices and heterogeneous communicating objects. Several IoT operating systems can be deployed on those nodes, including Contiki [79] and RIOT [73] [80].

## F. A Thought on RPL Deployments

Although it has a while since RPL standard was developed, we are yet to see real-world deployments of applications implementing the RPL protocol [17][20]. This was attributed in [17] to the complexity of the protocol related to its large set of features, that renders the protocol a very resource-intensive for constrained devices. In our view, there are two primary reasons for not seeing such wide deployments of RPL. First, RPL still suffers from many issues related to its scalability especially in bi-directional large-scale networks. This include the inefficiency of both modes of operations specified for *downward* routes and the under-specification of mechanisms used to control the construction of such routes (e.g. the policy of DAO emission) [16]. Second, the emergence of new proprietary lower power long range wireless technologies that cast doubts on the feasibility of using multi-hop routing in RPL-based LLN applications such as Wi-Fi HaLow [49] and LoRaWAN [134]. For instance, the trend in Libelium, one of the big players in the IoT market, is the use of star topology based on LoRaWAN and SigFox [135] technologies in its IoT applications, which may eliminate the need for multi-hop routing and consequently slow the adoption of RPL.

## VII. CONCLUSION

This article presents an extensive survey and in-depth comparative analysis of various routing enhancements made toward overcoming the weaknesses of the IPv6 Routing Protocol for Low-power and Lossy Networks (RPL). We started the discussion by first presenting an introductory background of the Low-power and Lossy Networks (LLNs), their communication technologies, their routing requirements based on RPL standard was specified conducting a detailed discussion of the standard known limitations in the light of these requirements. In particular, the survey aims at answering the question of whether RPL extensions have succeeded in overcoming its reported limitations related to the core operations of the standard (i.e. RPL's objective functions, routing maintenance and the downward routing primitives). We thoroughly overviews up-to-date research efforts that was proposed to address RPL's weaknesses and looked at the challenges and pitfalls they have faced. We found that, although a plenty of solutions have been introduced with the intent to boost further the efficiency of RPL, the majority of these solutions have serious pitfalls that undermine achieving the sought objectives and, thus many issues that were supposed to be solved by those extensions remain open for research. Among several others, these pitfalls include: the unrealistic operation conditions, the absence of real large-scale testbeds evaluations, the under-specification of metric-composition, and the greater complexity induced by some proposed solutions. In addition, we found that RPL has a serious scalability issue in bi-directional large-scale networks and neither of proposed solutions have efficiently address this concern, which might be the key reason undermining RPL large-scale deployments. We have also noticed that the emergence of lower power long-range wireless technologies may affect negatively the adoption of RPL and put into question the feasibility of multipath routing as a whole. In the light of this survey, we emphasized the need for further research efforts highlighting the main research directions, especially those hindering the adoption of the standard in large-scale deployments.

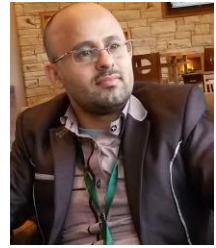

**Baraq Ghaleb** received the BSc degree in computer science from University of Jordan, Amman, Jordan in 2009 and the MSc degree from Jordan University of Science and Technology, Irbid, Jordan, in 2013. He is currently pursuing the Ph.D. degree in applied computing at Edinburgh Napier University, Edinburgh, UK. His current research interests include Routing protocols in Low-power and Lossy Networks and Internet of Things, Security of LLNs and IoT in addition to data mining. He holds one patent in the field of IoT Routing. He is a Student Member of the IEEE.

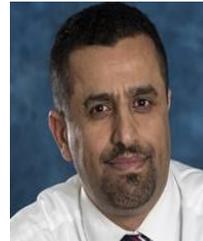

**Ahmed Y Al-Dubai** received the B.Sc. degree in computer science from Mutah University in 1996, the M.Sc. degree in computer science from Al al-Bayt University in 1999, and was awarded the Ph.D. in computing from the University of Glasgow in 2005. He is currently a Professor in the School of Computing where he leads the Networks Research Group at Edinburgh Napier University. His research interests span the areas of Pervasive and Mobile Computing, Communication Protocols and Algorithms. Intelligent Communication & Networks, and Internet of Things. He led several projects funded by the Scottish Funding Council, EU, Carnegie Trust, Royal Society, World Bank and other international funding bodies. He published widely in top-tier journals and leading IEEE/ACM international conferences. He has been the recipient of several international academic awards and recognitions, including Best Papers Awards at several international ACM/IEEE conferences. He is a member of several editorial boards of scholarly journals, and is Guest Editor for more than 20 Special Issues in International Journals. Moreover, he served as the chair/co-chair of more than 30 international conferences and workshops. He is a fellow of the British Higher Education Academy and a Senior Member of the IEEE.

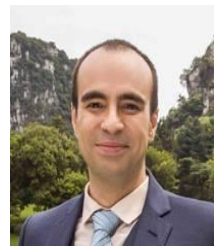

**Elias Ekonomou** is a Lecturer in the School of Computing at Edinburgh Napier University and a Security Architect at Sitekit. His research is centred around computer security, with a preference in lightweight applications and the contexts of IoT and WSN, where he completed his PhD in the University of Salford (2011). He is also has a strong interest in cloud-based federated identity, e-Health and social care.

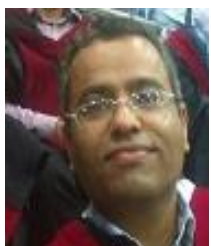

**Ayoub Alsarhan** received his Ph.D. degree in Electrical and Computer Engineering from Concordia University, Canada in 2011, his M.Sc. Degree in Computer Science from Al-Bayt University, Jordan in 2001, and B.E. degree in Computer Science from the Yarmouk University, Jordan in 1997. He is currently an Associate Professor at the Computer Information System Department of the Hashemite University, Zarqa, Jordan. His research interests include Cognitive Networks, Parallel Processing, Cloud Computing, Machine Learning, and Real Time Multimedia Communication over Internet.

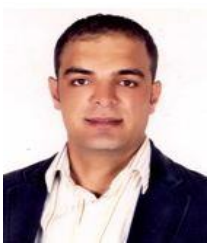

**Youssef Nasser** obtained his Ph.D. from the National Polytechnic Institute of Grenoble in 2006 and HDR in 2015. In 2003, he joined the Laboratory of Electronics and Information Technologies in Grenoble where he worked on B3G and 4G systems. In 2007, he has been working as a research engineer with INSA-Rennes and then joined the American University of Beirut (AUB) in 2010. His interests concern RF impairments, 4G and 5G systems, broadcasting, and localization. He has been involved in several research EU projects. He published more than 120 papers in top-tier journals and conferences.

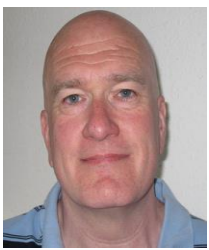

**Lewis M. Mackenzie** graduated with a B.Sc. in mathematics and natural philosophy from the University of Glasgow, U.K., in 1980. He was awarded the PhD in 1984, also from the University of Glasgow, for his work in the development of multicomputers for use in nuclear physics. He now lectures at the Department of Computing Science at the University of Glasgow, which he joined in 1985. His current research interests include multicomputers, high-performance networks, mobile ad hoc and vehicular networks, IoT and hypercomputation. He is a member of the IEEE.

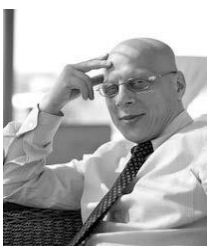

**Azzedine Boukerche** (FIEEE, FEiC, FCAE, and FAAAS) is a Distinguished University Professor and holds a Canada Research Chair Tier-1 position with the University of Ottawa. He is founding director of the PARADISE Research Laboratory and the DIVA Strategic Research Centre, University of Ottawa. He has received the C. Gotlieb Computer Medal Award, Ontario Distinguished Researcher Award, Premier of Ontario Research Excellence Award, G. S. Glinski Award for Excellence in Research, IEEE Computer Society Golden Core Award, IEEE CS Meritorious Award, IEEE TCPP Leaderships Award, IEEE ComSoc ASHN Leaderships and Contribution Award, and University of Ottawa Award for Excellence in Research. He serves as an associate editor for several IEEE transactions and ACM journals, and is a Steering Committee Chair for several IEEE and ACM international conferences. His current research interests include wireless ad hoc and sensor networks, wireless networking and mobile computing, wireless multimedia; QoS service provisioning, performance evaluation and modeling of large-scale distributed and mobile systems, and large scale distributed and parallel discrete event simulation. He has published extensively in these areas and received several best research paper awards for his work. He is a Fellow of IEEE, a Fellow of the Engineering Institute of Canada, the Canadian Academy of Engineering, and the American Association for the Advancement of Science.